\documentclass[pra,onecolumn,floatfix,a4paper,superscriptaddress]{revtex4}
\usepackage{bm,color,graphicx,amsmath,txfonts}
\usepackage{float}

\usepackage[colorlinks, citecolor=blue,linkcolor=blue]{hyperref}

\begin{document}

\title{Barnett effect boosted nonreciprocal entanglement and EPR-steering in magnomechanics in the presence of coherent feedback loop}

\author{Noura Chabar}
\affiliation{LPTHE-Department of Physics, Faculty of sciences, Ibnou Zohr University, Agadir, Morocco}	

\author{Mohamed Amazioug}  \thanks{m.amazioug@uiz.ac.ma}
\affiliation{LPTHE-Department of Physics, Faculty of sciences, Ibnou Zohr University, Agadir, Morocco}

\begin{abstract}{We propose an experimental scheme for enhancing entanglement, achieving asymmetric Einstein-Podolsky-Rosen (EPR) steering, and creating nonreciprocal quantum correlations within a hybrid system. This system integrates a yttrium iron garnet (YIG) sphere, which exhibits magnon-phonon coupling via magnetostriction, with a silica sphere featuring optomechanical whispering-gallery modes. By tuning the Barnett effect through the magnetic field direction, our system enables controllable asymmetric EPR steering and nonreciprocal entanglement between both directly and indirectly coupled modes. We demonstrate that adjusting the reflectivity of a beam splitter can boost stationary quantum steering and entanglement, effectively countering thermal noise. This approach allows for the generation of multipartite entanglement and both one-way and two-way steering. The proposed system is experimentally feasible and holds significant promise for various quantum information applications.}\\
\vspace{0.25cm}\textbf{Keywords}: Coherent feedback; Entanglement; Gaussian quantum steering; Nonreciprocity; Yttrium Iron Garnet (YIG); Barnett effect (BE).
\end{abstract}

\maketitle

\section{Introduction}

Generally, quantum entanglement is a special quantum phenomenon in the area of quantum information technology and science, such as quantum teleportation \cite{bennett1993teleporting,bouwmeester1997experimental,furusawa1998unconditional}, quantum metrology \cite{degen2017quantum,giovannetti2011advances}, quantum cryptography \cite{beveratos2002single}, quantum logic operations \cite{knill2001scheme}, and fundamental tests of quantum mechanics \cite{hensen2015loophole,giustina2015significant,shalm2015strong}, etc.
Additionaly to quantum entanglement, Einstein–Podolsky–Rosen (EPR) steering \cite{14a} as a subset of entanglement,  is a fundamental type of quantum correlation that lies between Bell nonlocality and entanglement in the hierarchy of nonclassical correlations \cite{15a,16a,17a}. Originally introduced by Schrödinger to highlight the nonlocality of EPR states \cite{18a,19a}. A key and unique characteristic of quantum steering is its inherent directionality \cite{20a}. Notably, steerable states in massive and macroscopic systems offer an avenue to probe foundational aspects of quantum mechanics and facilitate the implementation of quantum information protocols (QIP). Owing to rapid technological progress, recent research has increasingly focused on realizing these phenomena in optomechanical \cite{23a,24a}, magnomechanical systems \cite{26a} and magnonics system \cite{NR}. It is due to advancements in present-day technology that the most promising feature of quantum steering, one-way steering has been experimentally realized \cite{29a,30a}. Magnons have become a focus of quantum information research. This includes studies on their interaction with superconducting qubits \cite{15pla} and phonons \cite{16pla}. Additionally, magnons give rise to intriguing phenomena such as magnon-induced transparency \cite{17a,trns}, magnetically tunable slow light \cite{18pla}, and magnon blockade \cite{19pla} and  magnon squeezing \cite{squezz}.

Nonreciprocal entanglement was initially proposed in spinning resonators \cite{70b}, where rotation induces an asymmetric refractive index for clockwise and counterclockwise modes. This breaks time-reversal symmetry, enabling directional entanglement between photons and phonons while significantly suppressing it in the opposite direction. In addition to this mechanism, nonreciprocal entanglement has been explored through the Sagnac effect \cite{72b,73b}, magnon Kerr nonlinearity \cite{75b}, and chiral coupling \cite{77b}. However, the potential to realize nonreciprocal bipartite and tripartite entanglement by harnessing the Barnett effect within a cavity magnomechanical (CMM) system remains largely unexplored.

Opto-electromechanical systems \cite{bagci2014optical,andrews2014bidirectional,forsch2020microwave,zhong2020proposal,barzanjeh2011entangling,zhong24,Painter20,Jiang20,Rob24} provide a promising approach for generating microwave-optics entanglement by leveraging the properties of mechanical modes, which can interact with and couple to all electromagnetic fields. This can be achieved by simultaneously activating the optomechanical Pdc and beamsplitter interactions. Alternatively, it can be realized by enabling the optomechanical  Pdc and beamsplitter (Bs) interactions while utilizing the mechanical oscillator as an intermediary to distribute the quantum correlations generated in the parametric down-conversion (Pdc) process. Another approach involves directly coupling an electro-optic system and activating the Pdc interaction \cite{tsang2010cavity,rueda2019electro}, a method recently demonstrated \cite{sahu2023entangling}. Additionally, other strategies have been proposed, such as employing a cavity optomagnomechanical system \cite{fan2023microwave}. 
Another recent  experimently technique is by using a hybrid opto-magnomechanical system \cite{bs}. In this setup, a yttrium iron garnet (YIG) sphere is placed in physical contact with a silica microsphere, and the combined structure is embedded within a microwave cavity.  In our approach we propose a hybrid opto-magnomechanical system the system is composed of one magnon and one  silica samples placed inside a microwave cavity (MC). The two samples are connected physically in the MC. The YIG sphere hosts a magnon mode and a mechanical vibration mode, which are coupled through the magnetostrictive force \cite{Zuo24}. Meanwhile, the silica sample supports an optical whispering-gallery mode (WGM) and a mechanical vibration mode, interacting via the optomechanical coupling \cite{aspelmeyer2014cavity}. The two mechanical modes are coupled through direct physical contact between the two samples \cite{shen22}, while the magnon mode interacts with a microwave cavity mode via magnetic dipole coupling. The other nondirectly coupled modes are entangled as a result of the various direct couplings in the system. Importantly, a strong coupling between magnon modes and microwave cavity photons has been experimentally demonstrated, establishing cavity magnomechanical systems as a promising platform for continuous-variable quantum information processing \cite{exp22, exp23}

The uses of Coherent feedback (CF) has recently gained significant attention in  QIP . Feedback control has attracted growing interest and has made significant contributions to both theoretical and experimental research \cite{exp60,exp61,pr1}. By coherently routing a system's output back into its input through a beam splitter, this approach enables effective control while avoiding the disruptive effects of measurements. Furthermore, CF has been utilized for a range of applications, including noise suppression \cite{23}, squeezing \cite{24}, quantum state transfer \cite{26}, qubit coherence protection \cite{27}, entanglement enhancement \cite{28}, and transparency \cite{29}. 
The Barnett effect (BE) \cite{44, 47} describes the magnetization phenomenon that occurs when a magnetic material, such as a YIG sample, undergoes rapid rotation, causing a shift in the Barnett frequency, which can be effectively tuned to be either positive or negative by changing the direction of the magnetic field. This effect is rooted in the principle of angular momentum conservation: as the material rotates, the spin angular momentum of the electrons within it adjusts to preserve the total angular momentum. This adjustment results in the spontaneous magnetization of the material. The BE is a fundamental physical phenomenon that provides insight into the magnetization behavior of rotating magnetic materials. The BE has now been a powerful tool to control the nonreciprocity of many quantum physics phenomena, such as controlling bipartite and tripartite entanglement in a magnomechanics system \cite{BE} and controlling nonreciprocal unconventional magnon blockade \cite{BE1}. Beyond its significance in basic physics research, it also has potential applications in various technological fields.

In this work, we present a new approach to operate a hybrid opto-magnomechanical system incorporating feedback control toll and the BE, with the aim of demonstrating the feasibility of achieving controllable EPR steering, enhancing the degree of multiparite entanglement, and controlling the nonreciprocal entanglement. We show that by fine-tuning the reflectivity parameter and the Barnett shift, we can enhance the generating stationary entanglement and control the asymmetric EPR steering. Our protocol is compatible with current technology and holds strong potential as a novel approach for generating multipartite entanglement and one-way steering.  We have successfully generated a multipartite entanglement and EPR steering by adopting both CF and BE. The presence of the BE makes quantum correlations more resilient toward temperature. The control of quantum correlations is enhanced by controlling both the parameter related to the CF and the Barnett shift. We have shown that the presence of the CF and BE is indispensable for the generation of quantum correlation between the optical mode and the phonon mode (generated by the magnon) and between the magnon mode and the optical mode. We show the role of the Barnett shift in generating a nonreciprocal bipartite entanglement.

The remainder of the paper is structured as follows. In Sec.~\ref{model}, we introduce the system under investigation, which consists of two coupled opto- and magnomechanical microspheres along with a microwave cavity. We present the Hamiltonian and Langevin equations for the system, demonstrate how to linearize the dynamics around the steady-state averages, and derive the covariance matrix in terms of quantum fluctuations, from which the entanglement is calculated. In Sec.~\ref{resul}, we explore the numerical results obtained after measures of QCs and the nonreciprocal entanglement. Finally, we conclude and summarize the main findings in Sec.~\ref{conc}.
		
\section{Model and equations}  \label{model}

\begin{figure}[h]
 	\centering
 	{\label{figure4a}\includegraphics[scale=0.35]{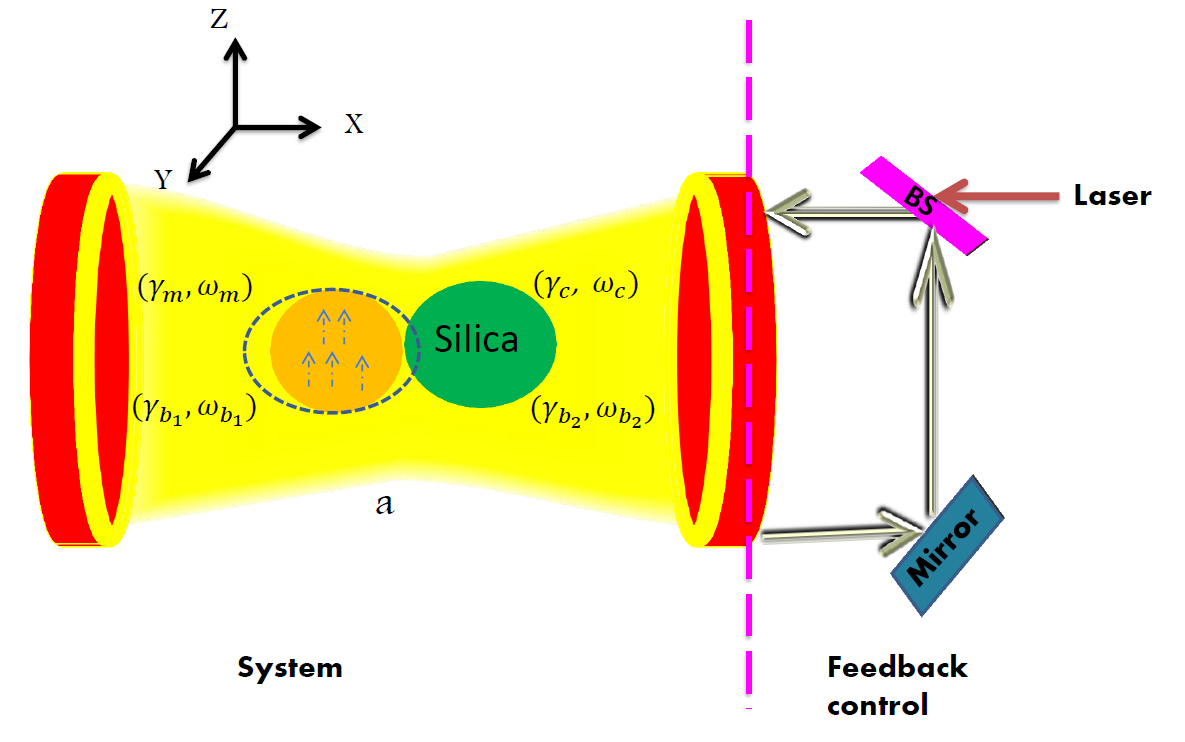}}
 	\put(-135,-1){(a)}
 	\hfil
 	{\label{figure4a}\includegraphics[scale=0.4]{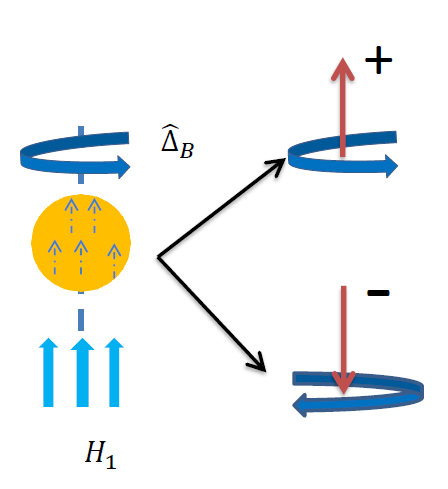}}
 	\put(-68,-1){(b)}
 	\hfil
 	{\label{figure4a}\includegraphics[scale=0.4]{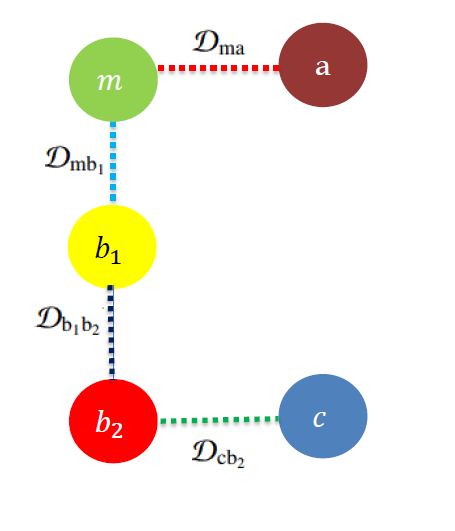}}
 	\put(-50,-3){(c)}
 	\caption{ (a) The hybrid system is composed of a YIG sample and a silica sample physically connected. The YIG sphere contains two modes: a magnon mode $\rm{m}$ and a vibration mode $\rm{b}_1$, which is driven by magnetostriction.  These modes have associated dissipation rates  $\gamma_{\rm{m}}$ and $\gamma_{\rm{b}_{1}}$, and operate at frequencies $\omega_m$, and $\omega_{\rm{b}_{1}}$, respectively. The silica sphere contains an optical whispering-gallery mode $\rm{c}$ and a mechanical mode $\rm{b}_2$, which are coupled through an optomechanical interaction. $\gamma_{\rm{c}}$ and $\gamma_{\rm{b}_{2}}$ are the dissipation rates associated with these modes and  $\omega_m$, and $\omega_{\rm{b}_{1}}$ are her frequencies, respectively. The beam splitter (BS) acts both as a feedback control element and as an input–output coupling device. (b) The insets depict the rotation directions of YIG  which gives rise to the angular frequencies $\pm\hat{\Delta}_{B}$. A bias magnetic field $H_1$ is applied to fully magnetize the YIG sphere. (c) Diagram illustrating the couplings between the various modes of the system via effective Pdc and BS interactions.
 	 }
 	\label{fig1}
 \end{figure}

The system under consideration, as depicted in Fig.~\ref{fig1}, is based on an experimentally feasible system \cite{bs}. Our system is composed of YIG and silica samples \cite{shen22}, which are physically connected and enclosed within a microwave cavity. The magnon mode (the Kittel mode) \cite{kittel1948theory} arises from the collective spin wave motion of a large number of spins within the YIG sphere. The excitation of the YIG samples is realized by subjecting the YIG to a uniform bias magnetic field and applying a microwave drive field. The magnon mode induced in the YIG samples has the advantage of interacting with the magnetostriction-induced mechanical mode (at $10^1$ MHz) \cite{Zuo24,zhang2016cavity,li2018magnon,potts2021dynamical,shen2022mechanical}, due to the large size of the YIG sphere, e.g., a 200-$\mu$m-diameter sphere used in Ref.~\cite{shen22}. Also, the magnon mode is able to interact with an MC mode via the magnetic dipole (MD) interaction \cite{huebl2013high,tabuchi2014hybridizing,zhang2014strongly}.
 	The silica sphere sustains both an optical whispering-gallery mode (WGM) and a mechanical mode, which interact through radiation pressure or the photoelastic effect \cite{aspelmeyer2014cavity}. Due to the close proximity of the two microspheres, their localized mechanical modes are directly coupled. The sizes of the spheres are carefully chosen so that their mechanical modes have similar frequencies and exhibit a linear beamsplitter-type coupling~\cite{shen22}. The Hamiltonian describing this hybrid system is given by \cite{bs}
\begin{equation}
	\begin{aligned}
		H/\hbar = &\sum_{s=\rm{a}, \rm{c}, \rm{b}_1, \rm{b}_2} \omega_s\, s^{\dagger} s 
		+ (\omega_{\rm{m}} + {\Delta}_B) \rm{m}^{\dagger} \rm{m}
		+ {\mathcal{D}}_{\rm{m} \rm{a}} (\rm{m}^{\dagger} \rm{a} + \rm{a}^{\dagger} \rm{m}) \\
		&+ {\mathcal{D}}_{\rm{m} \rm{b}_1} \rm{m}^{\dagger} \rm{m} (\rm{b}_1 + \rm{b}_1^{\dagger})
		- {\mathcal{D}}_{\rm{c} \rm{b}_2} \rm{c}^{\dagger} \rm{c} (\rm{b}_2 + \rm{b}_2^{\dagger}) 
		+{\mathcal{D}}_{\rm{b}_1 \rm{b}_2} (\rm{b}_1^{\dagger} \rm{b}_2 + \rm{b}_2^{\dagger} \rm{b}_1) \\
		&+  i \psi \mathcal{E} (\rm{c}^{\dagger} e^{-i \omega_{n_2} t} - \rm{c} e^{i \omega_{n_2} t})
			+ i \Omega (\rm{m}^{\dagger} e^{-i \omega_{n_1} t} - \rm{m} e^{i \omega_{n_1} t}),
	\end{aligned}
\end{equation}

the operators \( s = \rm{a}, \rm{m}, \rm{c}, \rm{b}_1, \rm{b}_2 \) (with \( s^\dagger \) denoting their respective creation operators) represent the annihilation operators of the microwave cavity mode, the magnon mode, the optical cavity mode, and the two mechanical modes, respectively. These operators satisfy the canonical commutation relation \( [s, s^\dagger] = 1 \). The corresponding resonant frequencies are denoted by \( \omega_s \), while the coupling strengths  \( \mathcal{{D}}_{s_{1}s_{2}}  \)  (${s_{1}s_{2}}= \rm{m}\rm{a}, \rm{b}_1 \rm{b}_2, \rm{m}\rm{b}_1, \rm{c}\rm{b}_2  $ ) are the magnon-cavity interaction, the mechanical coupling between the two microsamples, and the bare magnomechanical (optomechanical) interactions, respectively. Notably, these coupling strengths can be significantly enhanced by strongly driving the magnon mode (optical cavity) using a microwave (laser) field. 
The system is driven by an external Hamiltonian given by $\frac{H_{0}}{\hbar} + \frac{H_{1}}{\hbar} = i \psi \mathcal{E} \left( \rm{c}^{\dagger} e^{-i \omega_{n_2} t} - \rm{c} e^{i \omega_{n_2} t} \right)
	+ i \Omega \left( \rm{m}^{\dagger} e^{-i \omega_{n_1} t} - \rm{m} e^{i \omega_{n_1} t} \right),$ where the first term represents a laser field driving the optical cavity mode , and  the second term describes a microwave field driving the magnon mode. The Rabi frequency $\Omega = \frac{\sqrt{5}}{4} \gamma \sqrt{N} H_p$, characterizes the coupling strength between the magnon mode and the driving magnetic field of frequency $ \omega_{n_{1}}$ and amplitude $ H_p$.  Here, \( \gamma \) is the gyromagnetic ratio, and \( N \) is the total number of spins in the YIG sphere. The rotation of the YIG sphere induces a magnetic field \( H' = \hat{\Delta}_ B / \gamma \), known as the Barnett field, where \( \gamma  \) is the gyromagnetic ratio. As a result, the magnon  frequency $\omega_{\rm{m}}$ shifts to $\omega_{\rm{m}} + \hat{\Delta}_B$ \cite{44zh}, where $\hat{\Delta}_B$ represents the modification induced in the magnon mode frequency. Additionally, a shift of \( \pm \hat{\Delta} _B \) occurs when the rotation of the YIG is aligned along the \( \pm z \) direction. The parameter   $\mathcal{E} =  \sqrt{\frac{2 \gamma_{\rm{c}} P_L}{\hbar \omega_{n_2}}} $ represent the coupling strength between the WGM and the driving laser field.  
 Here, \( P_L \) and \( \omega_{n_2} \) correspond to the power and frequency of the laser field, respectively, while \( \gamma_c \) denotes the decay rate of the WGM due to its coupling with the fiber.
   The beam splitter is characterized by real reflection and transmission coefficients, denoted as  ${\mathcal{L}} $ and $ \psi $ , respectively. These coefficients satisfy the relation \( \psi ^2 + {\mathcal{L}}^2 = 1 \), ensuring that no energy is absorbed within the beam splitter. According to its properties, the transmitted component of the input laser beam has an amplitude of \( \psi \mathcal{E} \), while the reflected component has an amplitude of \( -{\mathcal{L}} \mathcal{E} \). The transmitted beam portion  is then used to pump the  optical  cavity.
The above Hamiltonian leads to the following quantum Langevin equations (QLEs) by including the dissipations and input noises, which, in the interaction picture with respect to $\hbar\omega_{d_1}(\rm{a}^{\dagger}\rm{a}+\rm{m}^{\dagger}\rm{m})+\hbar\omega_{d_2}\rm{c}^{\dagger}\rm{c}$, are given by
\begin{align}
	\begin{aligned}
		\dot{\rm{m}} &= 
		-(i{\Delta}_{{{\rm{m}}}_{BF}} + \gamma_{\rm{m}}) \rm{m}
		- i \mathcal{D}_{\rm{m} \rm{b}_1} (\rm{b}_1 + \rm{b}_1^\dagger) \rm{m}
		- i \mathcal{D}_{\rm{m} \rm{a}} \rm{a}
		+ \Omega 
		+ \sqrt{2\gamma_{\rm{m}}} \rm{m}_{\mathrm{in}}, \\
		\dot{\rm{c}} &=
		-(i\Delta_{\rm{c}_{FB}} + \gamma_{\rm{c}_{FB}})\rm{c}
		+ i \mathcal{D}_{\rm{c} \rm{b}_2} (\rm{b}_2 +  \rm{b}_2^\dagger)\rm{c}
		+ \psi \mathcal{E}
		+ \sqrt{2\gamma_{\rm{c}}} \rm{c}^{\mathrm{in}}_{BF}, \\
		\dot{\rm{a}} &=
		-(i\Delta_{\rm{a}} + \gamma_{\rm{a}}) \rm{a}
		- i \mathcal{D}_{\rm{m} \rm{a}} \rm{m}
		+ \sqrt{2\gamma_{\rm{a}}} \rm{a}_{\mathrm{in}}, \\
		\dot{\rm{b}}_1 &=
		-(i\omega_{\rm{b}_1} + \gamma_{\rm{b}_1}) \rm{b}_1
		- i \mathcal{D}_{\rm{m} \rm{b}_1} \rm{m}^\dagger \rm{m}
		- i \mathcal{D}_{\rm{b}_1 \rm{b}_2} \rm{b}_2
		+  \sqrt{2\gamma_{\rm{b}_1}} \rm{b}_{1,\mathrm{in}}, \\
		\dot{\rm{b}}_2 &=
		-(i\omega_{\rm{b}_2} + \gamma_{\rm{b}_2}) \rm{b}_2
		+ i \mathcal{D}_{\rm{c} \rm{b}_2} \rm{c}^\dagger \rm{c}
		- i \mathcal{D}_{\rm{b}_1 \rm{b}_2} \rm{b}_1
		+ \sqrt{2\gamma_{\rm{b}_2}} \rm{b}_{2,\mathrm{in}},
	\end{aligned}
\end{align}

where $\Delta_{{{\rm{m}}}_{BF}}= \tilde{\Delta}_{{\rm{m}}} + \hat{\Delta}_{B}$ is the effective detuning due to the presence of the BE, the detunings of the corresponding modes are defined by  $\Delta_{\rm{a}} \,{=}\,\omega_{\rm{a}}-\omega_{d_1}$, $\Delta_{\rm{c}_{FB}}= \tilde{\Delta}_{\rm{c}}\ + 2 \gamma _{\rm{c}}  {\mathcal{L}} \sin \Theta$ (with  $\tilde{\Delta}_{\rm{c}}\,{=}\,\omega_{\rm{c}}-\omega_{d_2}$) and, $\gamma_s$ ($ s = \rm{a}, \rm{m}, \rm{c}, \rm{b}_1, \rm{b}_2$) is the dissipation rate of the corresponding mode. The FB toll also modifies the dissipation rate of the optical mode, where $\gamma_{\rm{c}_{FB}} = \gamma_{\rm{c}} (1 - 2 {\mathcal{L}} \cos \Theta),  $ is the dissipation rate after using the FB. The input noise operators for the optical mode are defined as follows 
 \begin{equation}
 	\begin{aligned}
 		& \left\langle c_{FB}^{\mathrm{in}}(t) \, c_{FB}^{\mathrm{in}\dagger}(t') \right\rangle = \psi^2 \left| 1 - \mathcal{L} e^{i\Theta} \right|^2 \left[ n_c(\omega_c) + 1 \right] \delta(t - t'), \\
 		& \left\langle c_{FB}^{\mathrm{in}\dagger}(t) \, c_{FB}^{\mathrm{in}}(t') \right\rangle = \psi^2 \left| 1 - \mathcal{L} e^{i\Theta} \right|^2 n_c(\omega_c) \delta(t - t'),
 	\end{aligned}
 \end{equation}
 where the parameter $\Theta$ is the phase of the output field, with $\rm{c}^{\mathrm{in}}_{FB} ={\mathcal{L}} e^{i \Theta} c^{out}+\psi c^{in}$, with $c^{out}$ being the output field.  The parameters $m_{in}$, $c^{in}$, $a_{in}$, $b_{1,{in}}$, and $b_{2,{in}}$ are the input noise operators, which are zero-mean and obey the following correlation functions: $\langle s_{in}(t)s_{in}^{\dagger}(t^{\prime})\rangle=\left[n_{s}(\omega_{s})+1\right]\delta(t-t^\prime)$, and $\langle s_{in}^\dagger(t)s_{in}(t^{\prime})\rangle=n_s(\omega_s)\delta(t-t^{\prime})$. The mean thermal excitation number of each mode $n_{s}(\omega_{s})=\left[\exp(\hbar\omega_{s}/k_{B}T)-1\right]^{-1}$,  with $T$ being the bath temperature. The generation of microwave-optics entanglement relies on sufficiently strong opto- and magnomechanical coupling, which play a crucial role in cooling the two low-frequency mechanical modes and generating magnomechanical entanglement. To achieve this, we apply two strong driving fields to the magnon and optical cavity modes, resulting in large steady-state amplitudes, \( |\langle \rm{m}\rangle|, |\langle \rm{c}\rangle| \gg 1 \). This allows us to linearize the system dynamics around these large mean values while neglecting small second-order fluctuation terms. Consequently, we obtain a set of linearized QLEs for the quantum fluctuations, which can be written using quadratures and in the matrix form of
\begin{align}
\dot{\hat{Q}}(t)= \hat{A} \hat{Q}(t)+ \hat{N}(t),
\end{align}
where $\hat{Q}=\big[\delta X_{\rm{b}_1}(t),\delta Y_{\rm{b}_1}(t),\delta X_{\rm{b}_2}(t),\delta Y_{\rm{b}_2}(t),\delta X_{\rm{m}}(t),  \delta Y_{\rm{m}}(t), $ $\delta X_{c}(t),\delta Y_{\rm{c}}(t),\delta X_{\rm{a}}(t),\delta Y_{\rm{a}}(t) \big]^{T}$ is the vector of the quadrature fluctuations, $n(t)=\Big[\! \sqrt{2\gamma_{\rm{b}_1}}X_{\hat{b}_1}^{in}(t),\sqrt{2\gamma_{\rm{b}_1}}Y_{\rm{b}_1}^{in}(t),\sqrt{2\gamma_{\rm{b}_2}}X_{\rm{b}_2}^{in}(t),$ $\sqrt{2\gamma_{\rm{b}_2}}Y_{\rm{b}_2}^{in}(t), \sqrt{2\gamma_{\rm{m}}}X_{\rm{m}}^{in}(t),\sqrt{2\gamma_{\rm{m}}}Y_{\rm{m}}^{in}(t),  \sqrt{2\gamma_{\rm{c}}}X_{\rm{c}}^{in}(t),\sqrt{2\gamma_{\rm{c}}}Y_{\rm{c}}^{in}(t), $ $\sqrt{2\gamma_{\rm{a}}}X_{\rm{a}}^{in}(t),  \sqrt{2\gamma_{\rm{a}}}Y_{\rm{a}}^{in}(t) \Big]^T$ is the vector of the input noises, and the quadratures are defined as $X_{s}=\frac{1}{\sqrt{2}}(s+s^{\dagger})$ and $Y_s=\frac{i}{\sqrt{2}}(s^\dagger-s)$, and $\delta X_s$ and $\delta Y_s$ are the corresponding fluctuations. Similarly, the associated input noise operators $X_s^{in}$ and $Y_j^{in}$ can be defined. The drift matrix $\hat{A}$ is given by
\begin{widetext}  
\begin{eqnarray}    
 \hat{A}=\begin{pmatrix}-\gamma_{\rm{b}_1}&\omega_{\rm{b}_1}&0&\mathcal{{D}}_{\rm{b}_1 \rm{b}_2}&0&0&0&0&0&0\\-\omega_{\rm{b}_1}&-\gamma_{\rm{b}_1}&-\mathcal{\hat{D}}_{\rm{b}_1 \rm{b}_2}&0&0&-\sqrt{2}\mathcal{\hat{D}}_{\rm{m}}&0&0&0&0\\0&\mathcal{{D}}_{\rm{b}_1 \rm{b}_2}&-\gamma_{\rm{b}_{2}}&\omega_{\rm{b}_2}&0&0&0&0&0&0\\-\mathcal{{D}}_{\rm{b}_1 \rm{b}_2}&0&-\omega_{\rm{b}_{2}}&-\gamma_{\rm{b}_{2}}&0&0&0&-\sqrt{2}\mathcal{\hat{D}}_{\rm{c}}&0&0\\\sqrt{2}\mathcal{\hat{D}}_{\rm{m}}&0&0&0&-\gamma_{\rm{m}}&{\Delta}_{{{\rm{m}}}_{BE}}&0&0&0&\mathcal{{D}}_{\rm{m}\rm{a}}\\0&0&0&0&-{\Delta}_{{{\rm{m}}}_{BE}}&-\gamma_{\rm{m}}&0&0&-\mathcal{{D}}_{\rm{m}\rm{a}}&0\\0&0&\sqrt{2}\mathcal{\hat{D}}_{\rm{c}}&0&0&0&-\gamma_{\rm{c}_{FB}}&\Delta_{\rm{c}_{FB}}&0&0\\0&0&0&0&0&0&-\Delta_{\rm{c} _{FB} }&-\gamma_{\rm{c}_{FB}}&0&0\\0&0&0&0&0&\mathcal{{D}}_{\rm{m}\rm{a}}&0&0&-\gamma_{\rm{a}}&\Delta_{\rm{a}}\\0&0&0&0&-\mathcal{{D}}_{\rm{m}\rm{a}}&0&0&0&-\Delta_{\rm{a}}&-\gamma_{\rm{a}}\end{pmatrix}.
\end{eqnarray}
\end{widetext}
For the system to reach its stable steady-state, it is necessary that all eigenvalues of the drift matrix $\hat{A}$ possess negative real parts \cite{stability}. We have defined the effective magno- and optomechanical coupling strength: $\mathcal{\hat{D}}_{\rm{m}}=-i\sqrt{2}{\mathcal{D}}_{{\rm{m}} {\rm{b}}_1}\langle {\rm{m}}\rangle$ and $\mathcal{\hat{D}}_{{\rm{c}}}=i\sqrt{2}{\mathcal{D}}_{{\rm{c}} {\rm{b}}_2}\langle {\rm{c}}\rangle$. The
steady-state averages of the magnon and optical modes are
\begin{align}
 \left<\rm{m}\right>=\frac{\Omega}{(i\tilde{\Delta}_{\rm{m}_{BF}}+\gamma_{\rm{m}})+\frac{\mathcal{{D}}_{\rm{m}\rm{a}}^{2}}{i\Delta_{\rm{a}}+\gamma_{\rm{a}}}}, \,\,\,\,\,  \left<\rm{c}\right>=\frac{\psi \mathcal{E}}{(i\tilde{\Delta}_{\rm{c}_{FB}}+\gamma_{\rm{c}_{FB}})},
 \end{align}
the effective detunings $\tilde{\Delta}_{{{\rm{m}}}_{BF}}=\Delta_{{\rm{m}}_{BF}}+2{\mathcal{D}} _{{\rm{m}} b_1} {\rm Re} \langle \rm{b}_{1} \rangle$ and $\tilde{\Delta}_{\rm{c}_{FB}}=\Delta_{\rm{c}_{FB}}-2g_{\rm{c} \rm{b}_2} {\rm Re} \langle \rm{b}_{2}\rangle$, which include the frequency shift due to the mechanical displacement jointly caused by the photo- and magnetoelastic interactions.  The steady-state averages of the mechanical modes are 
 \begin{align}
 	\left<\rm{b}_{1}\right> &= \frac{\left|\left<\rm{c}\right>\right|^{2} {\mathcal{D}}_{\rm{c} \rm{b}_2} {\mathcal{D}}_{\rm{b}_1 \rm{b}_2}
 		- \left|\left<\rm{m}\right>\right|^{2} {\mathcal{D}}_{\rm{m} \rm{b}_1}(i\gamma_{\rm{b}_2} - \omega_{\rm{b}_2})}
 	{{\mathcal{D}}_{\rm{b}_1 \rm{b}_2}^{2} - (i\gamma_{\rm{b}_1} - \omega_{\rm{b}_1})(i\gamma_{\rm{b}_2} - \omega_{\rm{b}_2})}, \label{eq:l1} \\
 	\left<\rm{b}_{2}\right> &= \frac{\left|\left<\rm{c}\right>\right|^{2} {\mathcal{D}}_{\rm{c} \rm{b}_2}(i\gamma_{\rm{b}_1} - \omega_{\rm{b}_1})
 		- \left|\left<\rm{m}\right>\right|^{2} {\mathcal{D}}_{\rm{m} \rm{b}_1} {\mathcal{D}}_{\rm{b}_1 \rm{b}_2}}
 	{{\mathcal{D}}_{\rm{b}_1 \rm{b}_2}^{2} - (i\gamma_{\rm{b}_1} - \omega_{\rm{b}_1})(i\gamma_{\rm{b}_2} - \omega_{\rm{b}_2})}. \label{eq:l2}
 \end{align}
 
 We note that the above drift matrix $\hat{A}$ is derived under the optimal conditions for the microwave-optics entanglement, i.e., $|\Delta_{\rm{a}}|,\, |\tilde{\Delta}_{{{\rm{m}}}_{BF}}|,\, |\tilde{\Delta}_{\rm{c}_{FB}}| \,\,{\simeq}\,\, \omega_{\rm{b}_{1}} \,\,{\simeq}\,\, \omega_{\rm{b}_{2}} \,\,{\gg}\, \, \gamma_s$, $s=\rm{a},\rm{m},\rm{c}$. These lead to the following approximate expressions: $\langle \rm{m}\rangle\simeq -i\Omega/(\tilde{\Delta}_{{{\rm{m}}}_{BF}}-\mathcal{{D}}_{\rm{m}\rm{a}}^{2}/\Delta_{\rm{a}})$, and $\langle \rm{c}\rangle\simeq-i\psi \mathcal{E} /\tilde{\Delta}_{\rm{c}_{FB}}$, which are pure imaginary numbers, and therefore the effective couplings $\hat{\mathcal{D}}_{\rm{m}}$ and $\hat{\mathcal{D}}_{\rm{c}}$ are approximately real. Due to the linearized dynamics and the Gaussian nature of the input noises, the steady state of the system is a five-mode Gaussian state, which can be characterized by a 10 $\times$ 10 covariance matrix (CM) $\hat{V}$, of which the entries are defined as $\hat{V}_{ij}=\langle \hat{Q}_i(t)\hat{Q}_j(t)+\hat{Q}_j(t)\hat{Q}_i(t)\rangle/2$ $(i,j=1,2,...,10)$. The steady-state CM can be obtained by directly solving the Lyapunov equation \cite{vitali2007optomechanical}
\begin{align}
\hat{A}\hat{V}+\hat{V}\hat{A}^T+\hat{D}=0,
 \end{align}
where $\hat{D}=\mathrm{diag}\big[\gamma_{\rm{b}_1}(2n_{\rm{b}_{1}}+1),\gamma_{\rm{b}_{1}}(2n_{\rm{b}_{1}}+1),\gamma_{\rm{b}_2}(2n_{\rm{b}_2}+1),$
$\gamma_{\rm{b}_2}(2n_{\rm{b}_2}+1),\gamma_{\rm{m}}(2n_{\rm{m}}+1),\gamma_{\rm{m}}(2n_{\rm{m}}+1),\gamma_{\rm{c}} \psi^2\left|1-{\mathcal{L}} e^{i \Theta}\right|^2 (2n_{\rm{c}}+1),\gamma_{c} \psi^2\left|1-{\mathcal{L}} e^{i \Theta}\right|^2 (2n_{\rm{c}}+1),\gamma_{\rm{a}}(2n_{\rm{a}}+1),\gamma_{\rm{a}}(2n_{\rm{a}}+1)\big]$
 is the diffusion matrix  defined by $\hat{D}_{ij}\delta(t-t^{\prime})=\langle n_i(t)n_j(t^{\prime})+n_j(t^{\prime})n_i(t)\rangle/2$. We adopt the logarithmic negativity \cite{vidal2002computable,adesso2004extremal,plenio2005logarithmic} to quantify the different bipartite entanglement, which is defined as
\begin{align}
E_{ss^{\prime}}=\max[0,-\ln(2\eta^-)],
 \end{align}
where $\eta^{-}\equiv2^{-1/2}\left[\Sigma-(\Sigma^{2}-4\det \hat{V}_{4})^{1/2} \right]^{1/2}$, $\hat{V}_4$ is the 4 $\times$ 4 CM of the correspondant modes, which is in the form of
\begin{equation}
\hat{V}_4=\left[\hat{V}_s,\hat{V}_{ss^{\prime}};\hat{V}_{ss'}^{\rm T},\hat{V}_{s^{\prime}} \right], \quad s'\neq s.
\end{equation} 
 with $\hat{V}_{s^{\prime}}$, $\hat{V}_s$ and $\hat{V}_{ss^{\prime}}$ being the 2 $\times$ 2 blocks of $\hat{V}$, and $\Sigma\equiv\det \hat{V}_{s}+\det \hat{V}_{s^{\prime}}-2\det \hat{V}_{s^{\prime}}$, $\hat{V}_{s^{\prime}}$ and $\hat{V}_{s}$ are the submatrices corresponding to the reduced states of modes $s$ and $s^{\prime}$, respectively. Similarly, we can calculate the entanglement of any other two modes of the system.
To signify EPR steering, we adopt the computable measure of Gaussian steering proposed in Ref. \cite{64l} for arbitrary bipartite Gaussian states under Gaussian measurements. The steering in the direction from mode $s$ to mode $s^{\prime}\left(\mathcal{S}_{s \rightarrow s^{\prime}}\right)$ and in the opposite direction $\left(\mathcal{S}_{s^{\prime} \rightarrow s}\right)$ are quantified by
	$$
	\begin{aligned}
		\mathcal{S}_{s \rightarrow  s^{\prime}} & =\max \left\{0, E\left(2 \hat{V}_s  \right)-E(2 \hat{V})\right\},  \\
		\mathcal{S}_{s^{\prime} \rightarrow s} & =\max \left\{0, E\left(2 \hat{V}_{s^{\prime}}\right)-E(2 \hat{V})\right\},
	\end{aligned}
	$$
	
	where $E(\hat{V})=\frac{1}{2} \ln \operatorname{det} \sigma$ is the R\'enyi-2 entropy. $\mathcal{S}_{s  s^{\prime}}>0$ $\left(\mathcal{S}_{s^{\prime}  s}>0\right)$ implies that mode $s(s^{\prime})$ can steer mode $s^{\prime}(s)$ by gaussian measurements, and its value quantifies the degree of steering, i.e., a higher value of $\mathcal{S}$ indicates stronger Gaussian steerability. For studding the tripartite entanglement we refer to the minimum residual contangle.
\begin{equation}
\mathcal{R}_\tau^{\min } \equiv \min \left[\mathcal{R}_\tau^{m \mid a b}, \mathcal{R}_\tau^{b \mid a m}, \mathcal{R}_\tau^{a \mid m b}\right] 
\end{equation}
where $\quad \mathcal{R}_\tau^{i \mid j k} \equiv C_{i \mid j k}-C_{i \mid j}-C_{i \mid k} . \quad C_{i \mid j k}=\left(E^{i \mid j k}\right)^2 \quad$ is the contangle of subsystems, and $E^{i \mid j k}=\max \left[0,-\ln 2 \tilde{v}_{i \mid j k}\right]$, with $v_{i \mid j k}=\min \operatorname{eig}\left|\left[\oplus_{s=1}^3\left(-\sigma_y\right)\right] \mathcal{P}_{i \mid j k} \mathcal{V} \mathcal{P}_{i \mid j k}\right|$, where $\mathcal{P}_{a_{1} \mid a_{2} a_{3}}=$ $\sigma_z \oplus I \oplus I, \mathcal{P}_{a_{2} \mid a_{1} a_{3}}=I \oplus \sigma_z \oplus I$, and $\mathcal{P}_{a_{3} \mid a_{1} a_{2}}=I \oplus I \oplus \sigma_z$. When the  minimum residual contangle $\mathcal{R}_\tau^{\min }>0$  that mean   the presence of a genuine  tripartite entanglement.

\section{Results and analysis}  \label{resul}
 
We consider the hybrid setup illustrated in Fig.~\ref{fig1}. Our analysis focuses on exploring multipartite entanglement, genuine entanglement, quantum steering, and nonreciprocal entanglement among all the non-directly interacting modes in the system. The direct coupling between various modes in the system generates entanglement even between these non-directly coupled modes. We use the following experimentally feasible parameters \cite{42a, 44a}: $\omega_{\rm{a}, \rm{m}} / 2 \pi=10 \mathrm{GHz}, \omega_{\rm{b}_{1}} / 2 \pi=20.15 \mathrm{MHz}$, $\omega_{\rm{b}_{2}} / 2 \pi=20.11 \mathrm{MHz}$, optical cavity resonance wavelength $\lambda_c=1550 \mathrm{~nm}, \gamma_{\rm{a}, \rm{m}, \rm{c}} / 2 \pi=1 \mathrm{MHz}, \gamma_{{\rm{b}_{1}}, {\rm{b}_{2}}} / 2 \pi=100 \mathrm{~Hz}$, $\hat{\mathcal{D}}_{\rm{m}} / 2 \pi=0.7 \mathrm{MHz}, \hat{\mathcal{D}}_{\rm{c}}/ 2 \pi=2.7 \mathrm{MHz}$, and $T=10 \mathrm{mK}$.  We choose $\mathcal{D}_{\rm{ma}}/2\pi = 1.5\,\mathrm{MHz}$ and $\mathcal{D}_{\rm{b_1b_2}}/2\pi = 2.4\,\mathrm{MHz}$ as optimal couplings, we find the effective mean phonon numbers for the two mechanical modes to be $\bar{n}_{\rm{b_1}}^{\text{eff}} \simeq 0.11$ and $\bar{n}_{\rm{b_2}}^{\text{eff}} \simeq 0.08$. These values clearly indicate cooling to their quantum ground state. A relatively strong optomechanical coupling, $\hat{\mathcal{D}}_c/2\pi = 2.7\,\mathrm{MHz}$, is employed for cooling, a necessity given the hybrid system's multiple dissipation channels \cite{fan2023microwave}. This corresponds to a laser power of $P_L \cong 30\,\mathrm{mW}$ at $\mathcal{D}_{\rm{cb_2}}/2\pi = 100\,\mathrm{Hz}$ \cite{Zuo24}. A magnomechanical coupling of $\hat{\mathcal{D}}_{\rm{m}}/2\pi = 0.7\,\mathrm{MHz}$ is achieved with a drive power $P_0 \simeq 4\,\mathrm{mW}$ (corresponding to a drive magnetic field $H_d \simeq 3.3 \times 10^{-5}\,\mathrm{T}$) for $\mathcal{D}_{\rm{mb_1}}/2\pi = 0.1\,\mathrm{Hz}$ \cite{42a}. This power is determined using the relationship between the drive magnetic field $H_d$ and power $P_0$, given by $H_d=(1/R)\sqrt{(2P_0\mu_0/\pi c)}$ \cite{li2018magnon}. Here, $\mu_0$ represents the vacuum magnetic permeability, $c$ is the speed of electromagnetic waves in vacuum, and $R$ is the radius of the YIG sphere, set at $R=100\,\mathrm{\mu m}$. 

\begin{figure}[H]
\centering
\includegraphics[scale=0.27]{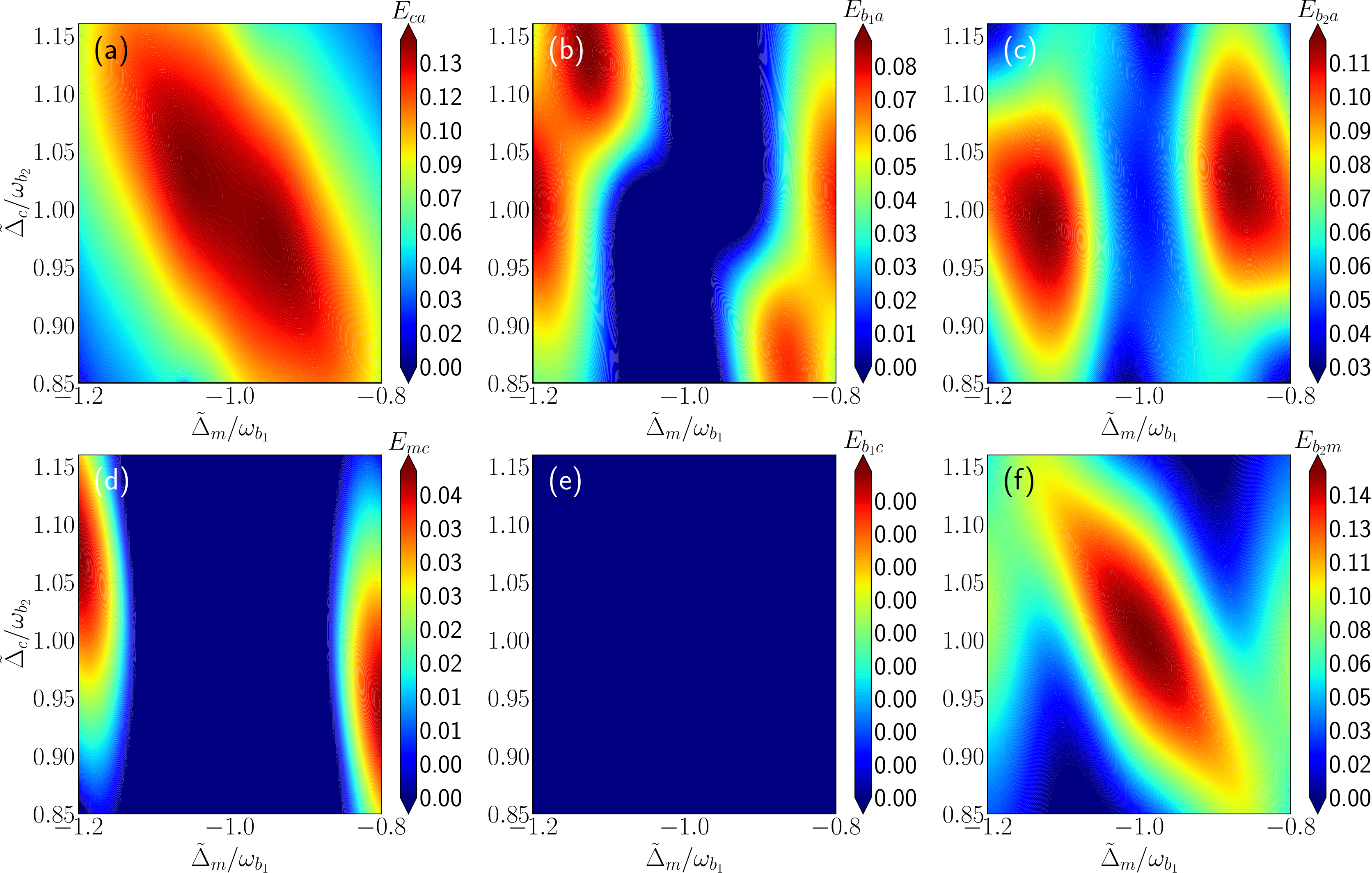}
\caption{Density plots of entanglement for various mode pairs: (a) $E_{\rm{ca}}$ (microwave-optical), (b) $E_{\rm{b_1c}}$ (mechanical $\rm{b_1}$-optical), (c) $E_{\rm{b_2a}} $ (mechanical $\rm{b_2}$-microwave), (d) $E_{\rm{mc}} $ (magnon-optical), (e) $E_{\rm{b_1c}} $ (mechanical $\rm{b_1}$-optical), and (f) $E_{\rm{b_2m}} $ (mechanical $\rm{b_2}$-magnon). All plots are shown as a function of the magnon detuning $\tilde{\Delta}_{\rm{m}}/\omega_{\rm{b_1}}$ and the optical detuning $\tilde{\Delta}_{\rm{c}}/\omega_{\rm{b_2}}$, with ${\mathcal{L}} =\hat{\Delta}_{B}=0$. Other parameters are provided in the text.}
\label{r}
\end{figure}
First, Fig.~\ref{r} presents density plots of entanglement for the  six non-directly coupled pairs, all as a function of the normalized magnon mode detuning $\tilde{\Delta}_{\rm{m}}$ and the optical mode detuning $\tilde{\Delta}_{\rm{c}}$. This is shown in the absence of the FB tool and Barnett shift ($\hat{\Delta}_B={\mathcal{L}}=0$). Figs.~\ref{r}(a) and (f) highlight the role of operating near the cooling regime of the mechanical mode (i.e., $\tilde{\Delta}_{\rm{m}}\simeq -\omega_{\rm{b_1}}$ and $\tilde{\Delta}_{\rm{c}}\simeq\omega_{\rm{b_2}}$) in generating significant optical-microwave entanglement and strong phonon $\rm{b_2}$-magnon entanglement. The microwave-optic entanglement reaches a maximum value of $\approx 0.13$. We also observe that efficient magnon-to-microwave state transfer is achieved when the microwave and magnon modes are nearly resonant ($\Delta_{\rm{a}} \simeq \tilde{\Delta}_{\rm{m}}$) and strongly coupled ($\mathcal{D}_{\rm{ma}} > \gamma_{\rm{m}}, \gamma_{\rm{a}}$)~\cite{yu20}. For the bipartitions $E_{\rm{b_1a}}$, $E_{\rm{b_2m}}$, and $E_{\rm{mc}}$, entanglement is maximized when $\Delta_{\rm{c}} \approx \omega_{\rm{b_1}}$ and  $\tilde{\Delta}_{\rm{m}}$ takes an appropriate value. However, we observe an absence of entanglement ($E_{\rm{b_1c}}=0$) between the mechanical mode $\rm{b_1}$ and the optical mode under the adopted numerical values.

\begin{figure}[H]
\centering
\includegraphics[scale=0.27]{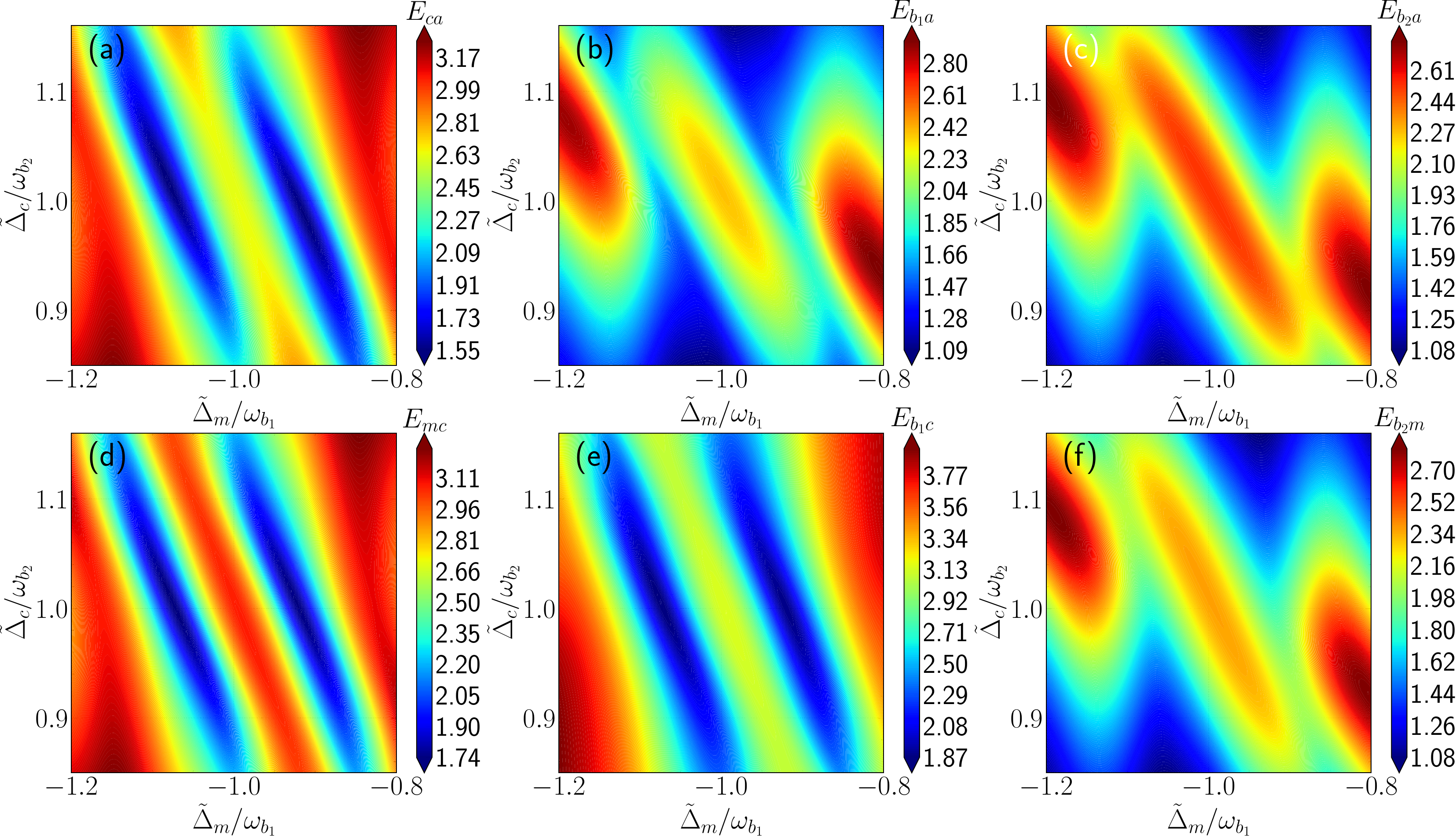}
\caption{Density plots of entanglement for various mode pairs: (a) $E_{\rm{ca}}$ (microwave-optical), (b) $E_{\rm{b_1c}}$ (mechanical $\rm{b_1}$-optical), (c) $E_{\rm{b_2a}} $ (mechanical $\rm{b_2}$-microwave), (d) $E_{\rm{mc}} $ (magnon-optical), (e) $E_{\rm{b_1c}} $ (mechanical $\rm{b_1}$-optical), and (f) $E_{\rm{b_2m}} $ (mechanical $\rm{b_2}$-magnon). Each plot shows entanglement as a function of magnon detuning $\tilde{\Delta}_{\rm{m}}/\omega_{\rm{b_1}}$ and optical detuning $\tilde{\Delta}_{\rm{c}}/\omega_{\rm{b_2}}$, with $\hat{\mathcal{L}} = 0.9$ and $\hat{\Delta}_{B}=0$. Other parameters are detailed in the text.}
\label{r1}
\end{figure}

\begin{figure}[h]
	\centering
	\includegraphics[scale=0.27]{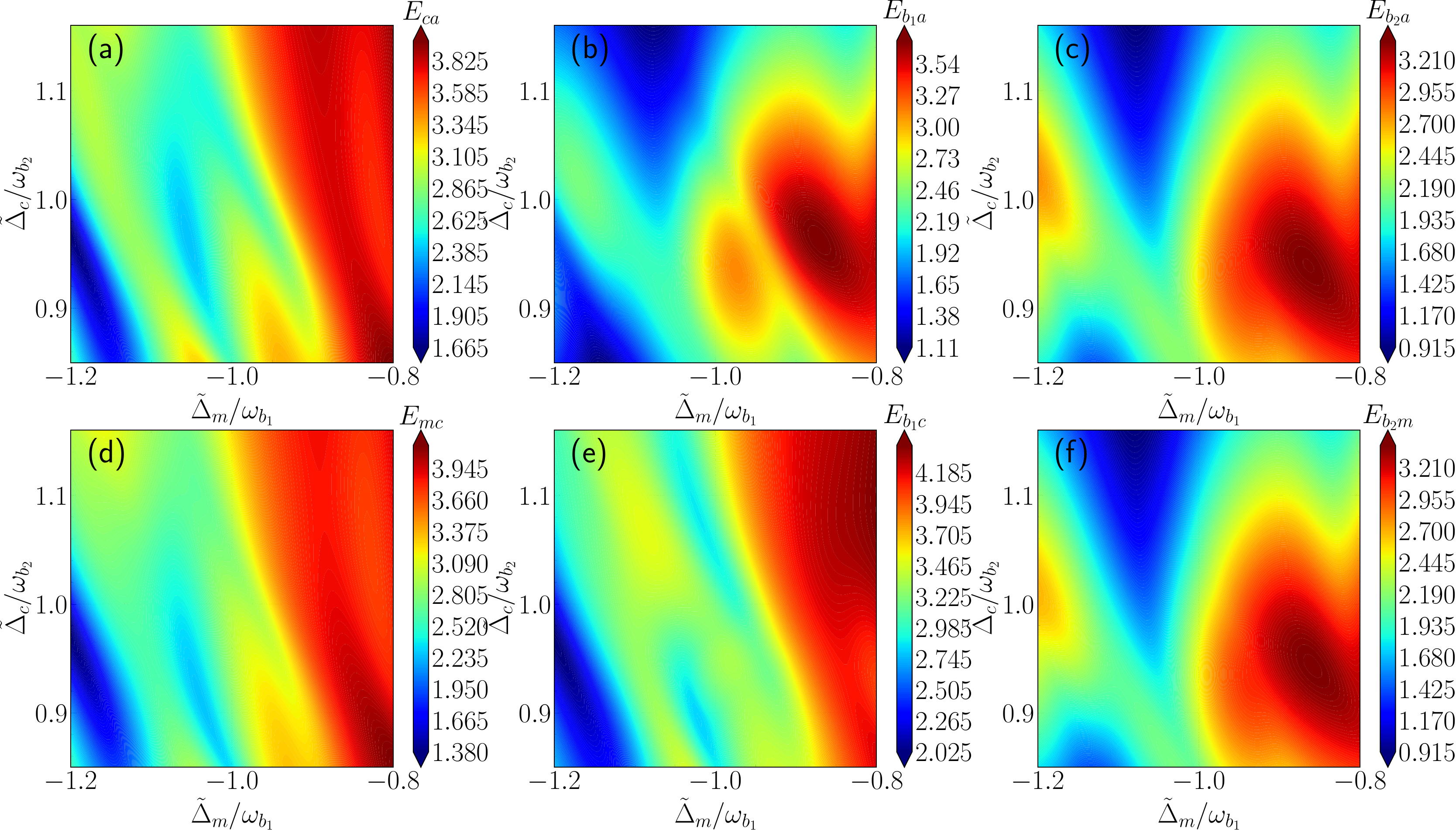}
	\caption{Density plots of entanglement for various mode pairs: (a) $E_{\rm{ca}}$ (microwave-optical), (b) $E_{\rm{b_1c}}$ (mechanical $\rm{b_1}$-optical), (c) $E_{\rm{b_2a}}$ (mechanical $\rm{b_2}$-microwave), (d) $E_{\rm{mc}}$ (magnon-optical), (e) $E_{\rm{b_1c}}$ (mechanical $\rm{b_1}$-optical), and (f) $E_{\rm{b_2m}}$ (mechanical $\rm{b_2}$-magnon). Each plot shows entanglement as a function of magnon detuning $\tilde{\Delta}_{\rm{m}}/\omega_{\rm{b_1}}$ and optical detuning $\tilde{\Delta}_{\rm{c}}/\omega_{\rm{b_2}}$, with $\mathcal{L} = 0.9$ and $\hat{\Delta}_{B}=0.2\,\omega_{\rm{b_1}}$. Other parameters are detailed in the text.}
	\label{r2}
\end{figure}

Next, we present the results when only FB is adopted (Fig.~\ref{r1}) and when both BE and FB are employed in the system (Fig.~\ref{r2}). The maximum optical-microwave entanglement in Fig.~\ref{r1}(a) (with $\hat{\Delta}_B = 0$ and $\mathcal{L}=0.2\,\omega_{\rm{b_1}}$) is approximately 3.8. This represents an enhancement compared to Fig.~\ref{r}(a), with further enhancement observed when combining FB with BE (Fig.~\ref{r2}(a), where $\hat{\Delta}_B =0.2\,\omega_{\rm{b_1}}$ and $\mathcal{L}=0.9$). The results clearly demonstrate the role of combining both BE and FB tools in significantly increasing entanglement. These observations and explanations also apply to the other bipartitions. To compare our findings with the case study in \cite{bs}, we select $\tilde{\Delta}_{\rm{c}}= \omega_{\rm{b_2}}$ and $\tilde{\Delta}_{\rm{m}}= - \omega_{\rm{b_1}}$ as common values that generate substantial entanglement between all studied modes. We observe that entanglement emerges between phonon mode $\rm{b_2}$ and magnon mode $\rm{m}$ when only the FB tool is used, and this entanglement is further enhanced when both FB and BE tools are employed.

\begin{figure}[H]
	\centering
	\includegraphics[scale=0.27]{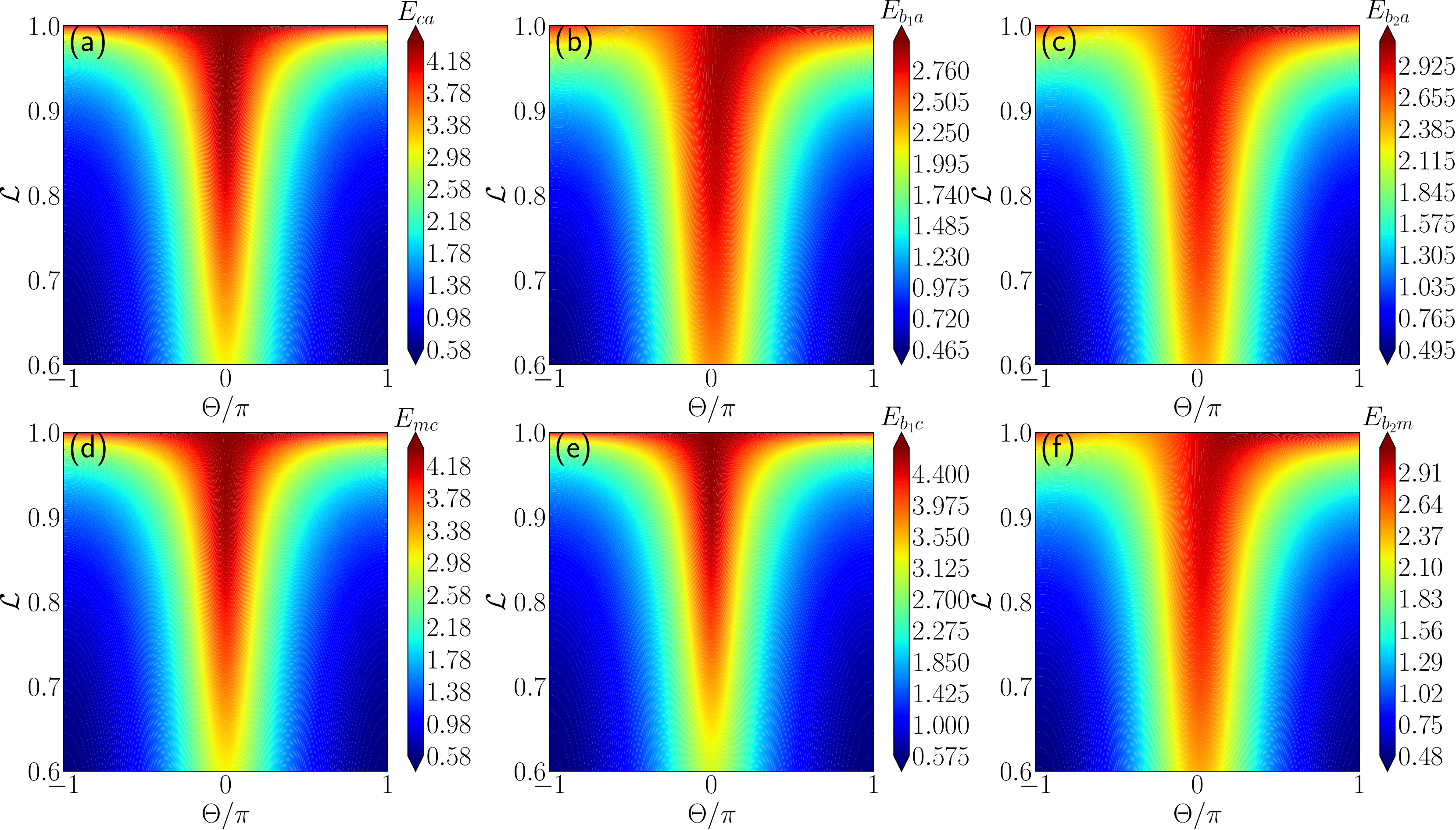}
	\caption{Density plots of entanglement for various mode pairs: (a) $E_{\rm{ca}}$ (microwave-optical), (b) $E_{\rm{b_1c}}$ (mechanical $\rm{b_1}$-optical), (c) $E_{\rm{b_2a}}$ (mechanical $\rm{b_2}$-microwave), (d) $E_{\rm{mc}}$ (magnon-optical), (e) $E_{\rm{b_1c}}$ (mechanical $\rm{b_1}$-optical), and (f) $E_{\rm{b_2m}}$ (mechanical $\rm{b_2}$-magnon). Each plot shows entanglement as a function of the phase shift $\Theta$ and the reflectivity parameter $\mathcal{L}$. All parameters are the same as those used in Fig.~\ref{r2}.}
	\label{r3}
\end{figure}

Then in Fig.~\ref{r3}, we plot the density of entanglement between the six modes as a function of the phase shift $\Theta$ and the reflectivity parameter $\mathcal{L}$. The entanglement is maximum near $\Theta=0$. We observe that entanglement increases monotonically with the increase of $\mathcal{L}$ when $\Theta$ is fixed at zero. In the following, we use $\Theta=0$ as an optimal value for generating entanglement.

\begin{figure}[H]
			\centering
			{\label{figure7a}\includegraphics[scale=0.27]{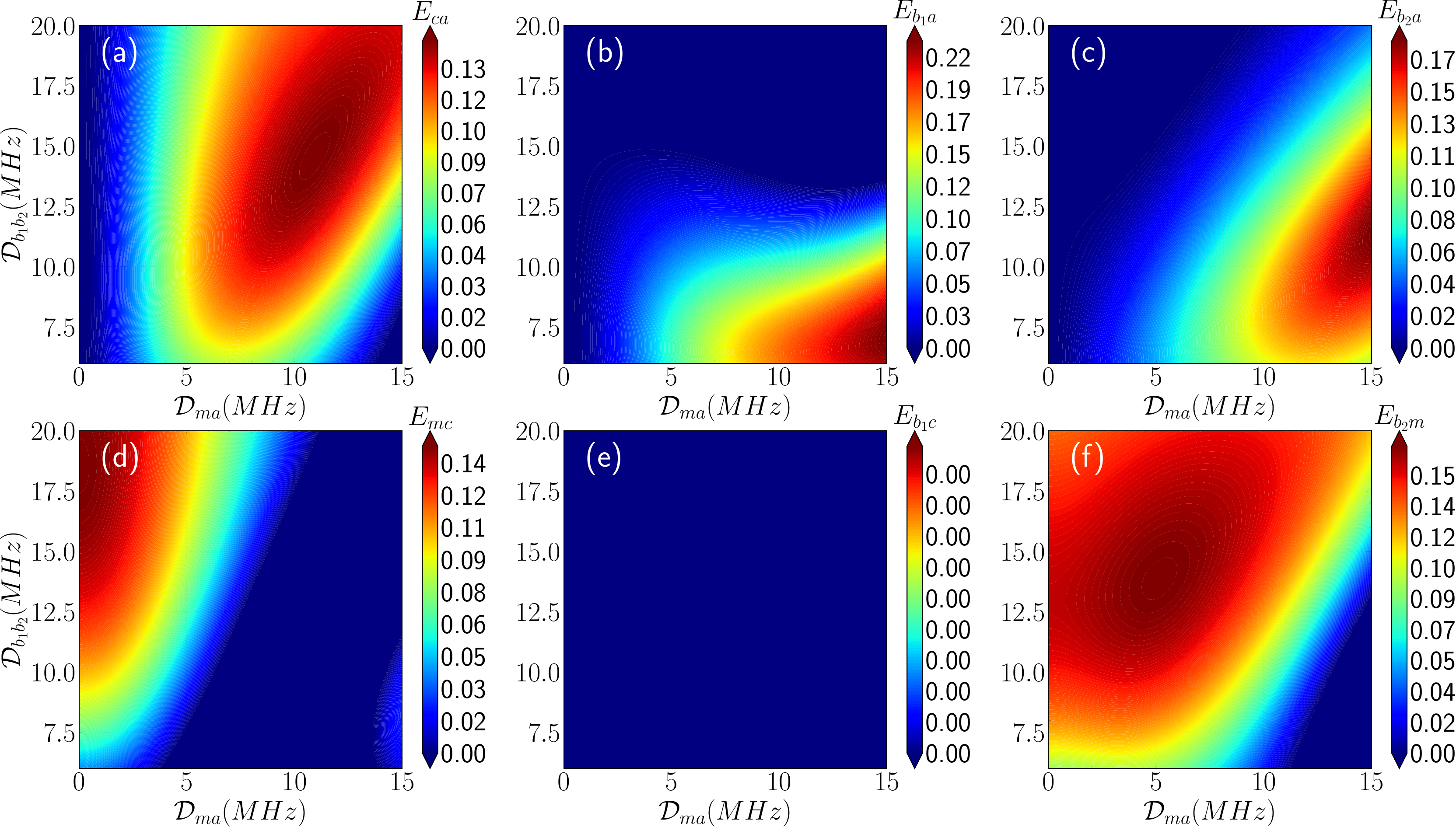}}			
			\caption{Density plots of entanglement for various mode pairs: (a) $E_{\rm{ca}}$ (microwave-optical), (b) $E_{\rm{b_1c}}$ (mechanical $\rm{b_1}$-optical), (c) $E_{\rm{b_2a}}$ (mechanical $\rm{b_2}$-microwave), (d) $E_{\rm{mc}}$ (magnon-optical), (e) $E_{\rm{b_1c}}$ (mechanical $\rm{b_1}$-optical), and (f) $E_{\rm{b_2m}}$ (mechanical $\rm{b_2}$-magnon). Each plot shows entanglement as a function of the microwave–magnon coupling $\mathcal{D}_{\rm{ma}}$ and the mechanical–mechanical coupling $\mathcal{D}_{\rm{b_1b_2}}$, with $\mathcal{L}=\hat{\Delta}_{B}=0$.}
			\label{GF25}
		\end{figure}

\begin{figure}[H]
			\centering
		{\label{figure7a}\includegraphics[scale=0.27]{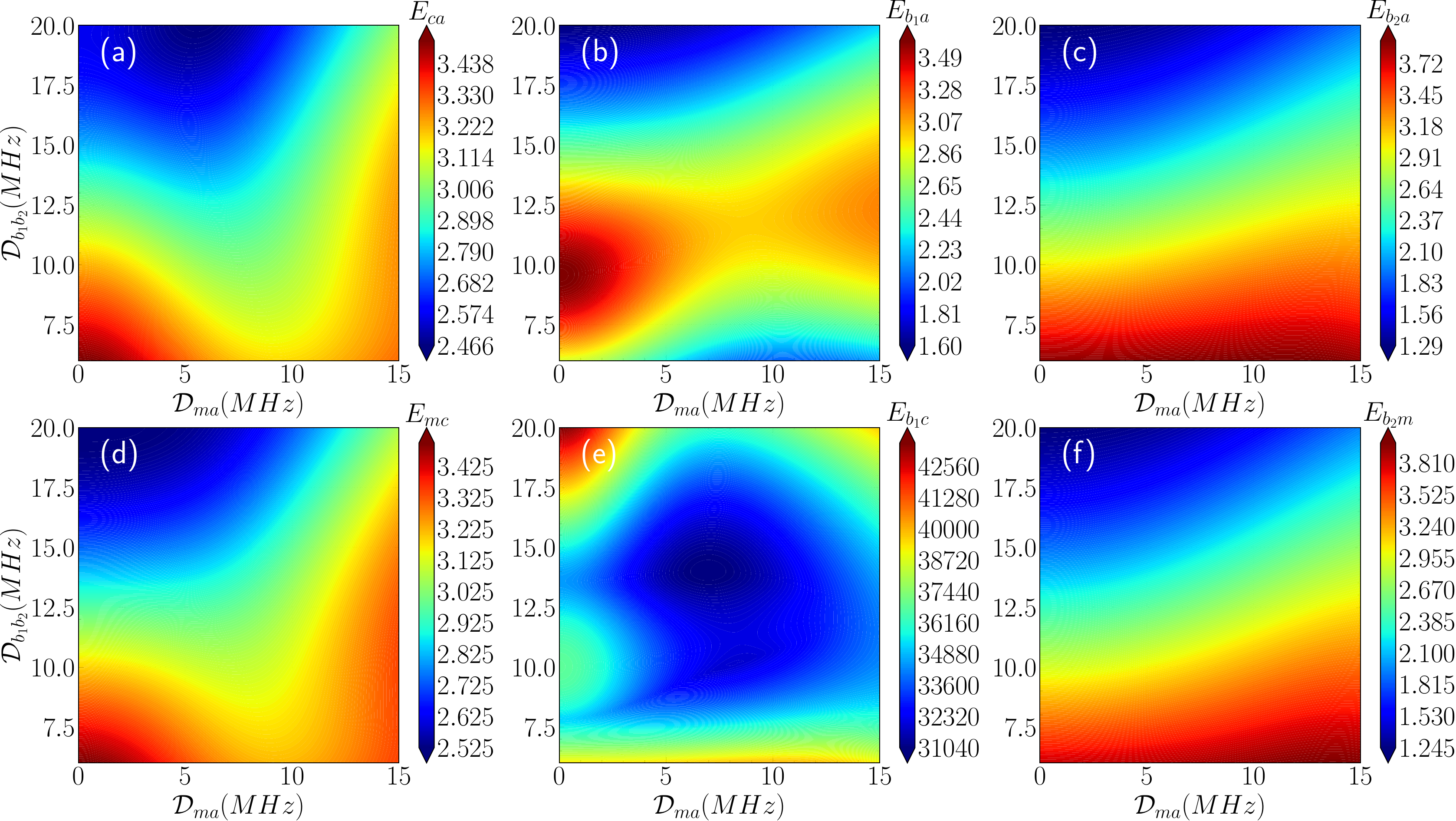}}\caption{Density plots of entanglement for various mode pairs: (a) $E_{\rm{ca}}$ (microwave-optical), (b) $E_{\rm{b_1c}}$ (mechanical $\rm{b_1}$-optical), (c) $E_{\rm{b_2a}}$ (mechanical $\rm{b_2}$-microwave), (d) $E_{\rm{mc}}$ (magnon-optical), (e) $E_{\rm{b_1c}}$ (mechanical $\rm{b_1}$-optical), and (f) $E_{\rm{b_2m}}$ (mechanical $\rm{b_2}$-magnon). Each plot shows entanglement as a function of the microwave–magnon coupling $\mathcal{D}_{\rm{ma}}$ and the mechanical–mechanical coupling $\mathcal{D}_{\rm{b_1b_2}}$, with $\mathcal{L}=0.9$ and $\hat{\Delta}_{B}=0.2\,\omega_{\rm{b_1}}$.}
			\label{GF2}
		\end{figure}

The steady-state entanglement density among the six investigated modes is presented in Figs.~\ref{GF25} and \ref{GF2}. These plots illustrate entanglement as a function of the microwave–magnon coupling $\mathcal{D}_{\rm{ma}}$ and the mechanical–mechanical coupling $\mathcal{D}_{\rm{b_1b_2}}$, under two distinct conditions: $\mathcal{L}=\hat{\Delta}_B=0$ (Fig.~\ref{GF25}) and $\hat{\mathcal{L}}=0.9$ with $\hat{\Delta}_B=0.2\,\omega_{\rm{b_1}}$ (Fig.~\ref{GF2}). A complementary behavior is clearly evident in the curves: microwave–optics entanglement arises from a partial transfer of entanglement from the magnon–optics entanglement, mediated by the microwave–magnon beamsplitter interaction.
		
Magnon–optics entanglement arises from the magnomechanical entanglement, being subsequently redistributed to the magnon–optics subsystem through the $\rm{b_1}$–$\rm{b_2}$–$\rm{c}$ (mechanical–mechanical–optical) pathway. This process is illustrated in Fig.~\ref{GF25} and \ref{GF2}. This mediates coupling between modes that are not directly interacting. Efficient quantum state transfer also requires mechanical modes $\rm{b_1}$ and $\rm{b_2}$ to have nearly resonant frequencies ($\omega_{\rm{b_1}} \simeq \omega_{\rm{b_2}}$) and be strongly coupled ($\mathcal{D}_{\rm{b_1b_2}}>\gamma_{\rm{b_1}}^{\text{eff}}, \gamma_{\rm{\hat{b}_2}}^{\text{eff}}$). Here, $\gamma_{\rm{\hat{b}_j}}^{\text{eff}} \gg \gamma_{\rm{\hat{b}}_j}(j=1,2)$ are the increased effective mechanical damping rates due to the optomechanical cooling interaction and the mechanical beamsplitter coupling. 
\begin{figure}[H]
	\centering
	{\label{figure5a}\includegraphics[scale=0.6]{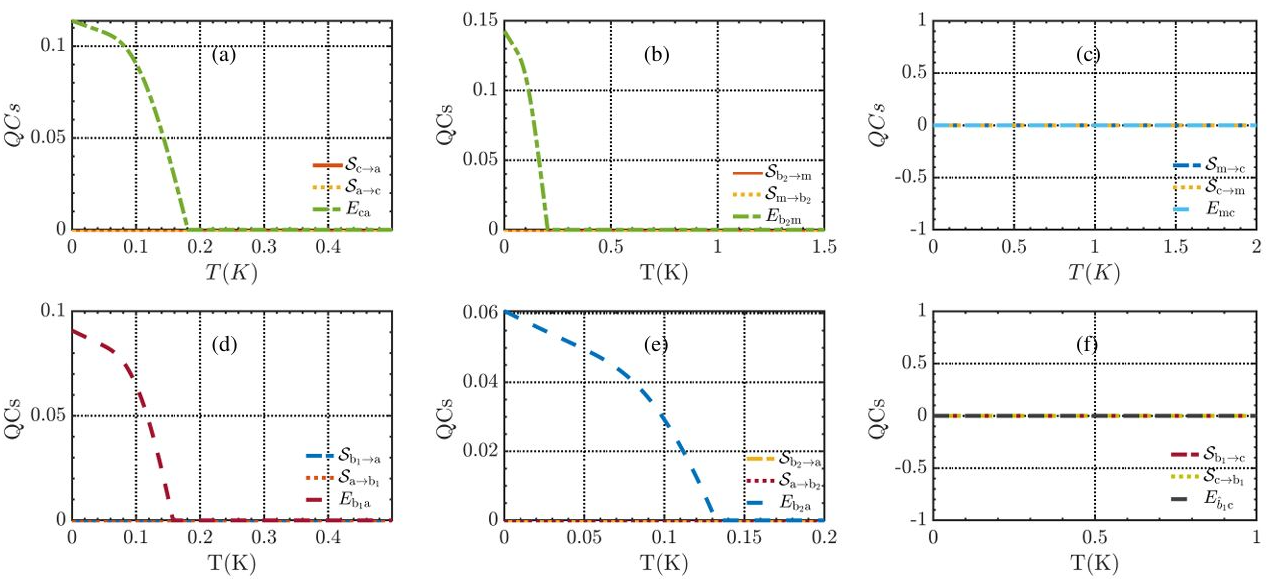}}	
	\caption{The Qcs (steerings $\mathcal{S}_{  {s } \rightarrow s'}$,  $\mathcal{S}_{ s' \rightarrow s}$ and entanglement $E_{ss' }$) between the six nondirectly  coupled modes, vary in respect to the temperature $T$ variation. For $ {\mathcal{L}}= \hat{\Delta}_{B}=0$. The other parameters are given in the text.}
	\label{tau0}
\end{figure}	
 	Next, we explore how both the BE and FB tools enhance entanglement and control asymmetric EPR-steering. Specifically, Fig. \ref{tau0}(a) shows the entanglement as a function of temperature $T$ when neither FB nor BE is into account ($\mathcal{L}=\hat{\Delta}_B=0$). In Fig. \ref{tau0}(a), we observe that the optical mode and microwave mode are not steerable, and their entanglement ($E_{\rm{ac}}$) quickly disappears after reaching a maximum value of $\approx 0.1$ at $T=0\,\mathrm{K}$. This behavior is consistent with the decoherence parameters ($\gamma_{\rm{a}}=\gamma_{\rm{c}}$) and low mechanical damping ($\gamma_{\rm{b_1,b_1}} \ll \gamma_{\rm{a,c}}$) used in this case, which are known factors for definitively non-steerable modes in any direction \cite{36sterr}. An absence of entanglement and steering is observed between the magnon-optics modes ($\mathcal{S}_{\rm{m} \rightarrow \rm{c}}= \mathcal{S}_{ \rm{c} \rightarrow \rm{m}}=E_{\rm{mc}}=0 $) and between the phonon($\rm{b_1}$)-optics modes ($\mathcal{S}_{\rm{b_1} \rightarrow \rm{c}}= \mathcal{S}_{ \rm{c} \rightarrow \rm{b_1}}=E_{\rm{b_1c}}=0 $). Furthermore, entanglement monotonically decreases with increasing temperature, a consequence of decoherence induced by the elevated thermal environment. When only the FB tool is applied (Fig.~\ref{P1}), entanglement among all the studied modes is enhanced. We also observe the emergence of entanglement between the magnon–optical and phonon $\rm{b_1}$–optical modes. Furthermore, the appearance of one-way and two-way quantum steering demonstrates our proposed system's capability in manipulating and controlling asymmetric steering. The initiation of steering between the optical and microwave modes can be attributed to the FB effect on changing the optical modes damping rate. The damping rate parameters of the two modes ($\rm{c}$ and $\rm{a}$) are unequal. This clearly shows that all steerable states are entangled; however, the converse isn't necessarily true, as entangled states aren't always steerable. This highlights that stronger quantum correlations (QCs) between modes are needed to achieve Gaussian steering than for entanglement alone. The resilience of entanglement to temperature variations differs across bipartitions, but it's clear that the temperature at which entanglement disappears is improved from the case in Fig.~\ref{tau0} to the case in Fig.~\ref{P}.  The insets of Figures \ref{tau0} and \ref{P} illustrate that tuning the reflectivity parameter $\mathcal{L}$ and the Barnett shift $\hat{\Delta}_{B}$ enhances the degree of entanglement at $T = 0 \mathrm{K}$.
 	
\begin{figure}[H]
 \centering
  {\label{figure44c}\includegraphics[scale=0.6]{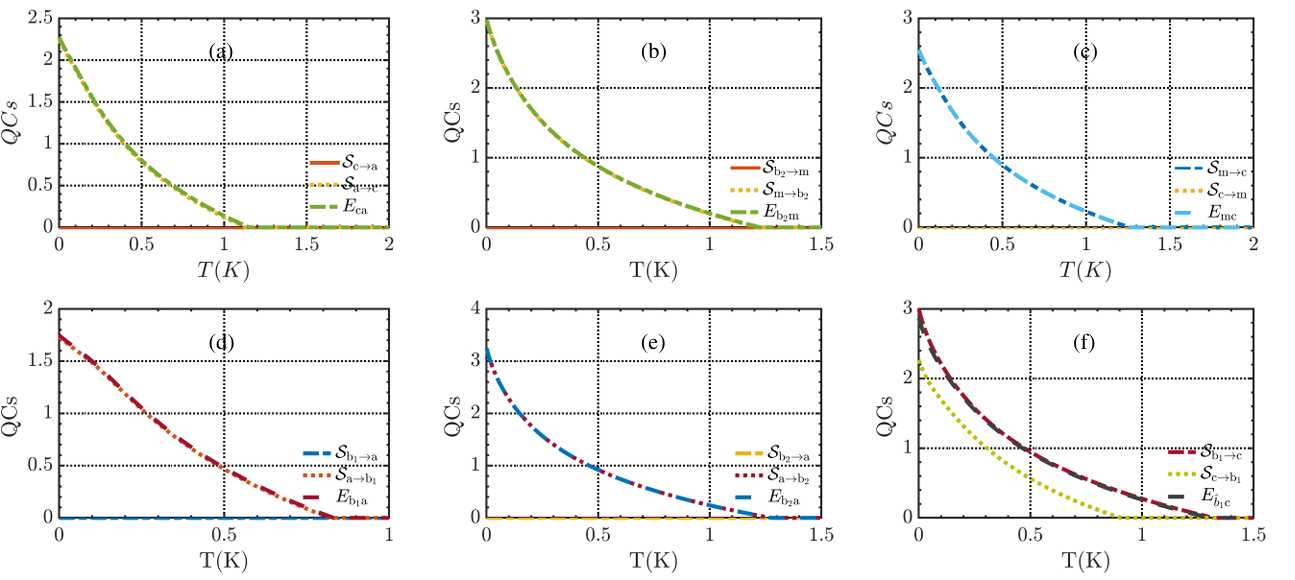}}
 \caption{ The Qcs (steerings $\mathcal{S}_{  {s } \rightarrow s'}$,  $\mathcal{S}_{ s' \rightarrow s}$ and entanglement $E_{ss' }$) between the six nondirectly  coupled modes, vary in respect to the temperature $T$ variation. For $ {\mathcal{L}}=0.9$ and $\hat{\Delta}_{B}=0$. }  
  			\label{P1}
  		\end{figure}
The Fig. \ref{P} clearly demonstrates that the simultaneous presence of both FB and BE significantly enhances the QCs across all bipartitions. For example, the microwave-optic entanglement reaches a peak value of approximately 2.23, representing a notable improvement over the result shown in Fig. \ref{tau0}(a). Moreover, the QCs among all bipartitions exhibit increased robustness against temperature-induced decoherence, as also illustrated in \cite{BE}.

Next, we discuss the effect of the Barnett shift ($\hat{\Delta}_{B}$) on the enhancement and nonreciprocity of entanglement between the non-directly coupled modes. Figure \ref{AF1} shows the entanglement as a function of temperature for different $\hat{\Delta}_{B}$ values, with parameters chosen to maintain the opto-magnonmechanical system within stable regions.

Fig. \ref{AF1} investigates the influence of both temperature and angular frequencies on the entanglement of various bipartitions, with a fixed reflectivity parameter $\mathcal{L}$. Our analysis reveals that entanglement can persist at higher temperatures when the YIG's rotation is orientated along a specific direction, indicating enhanced robustness against thermal noise under suitable parameter regimes. This effect is attributable to the Barnett shift. While bipartite entanglement generally strengthens as the temperature decreases, extremely low temperatures can lead to unwanted entanglement between other modes \cite{108bos}. By tuning the temperature $T$, an optimal level of bipartite entanglement can always be found. Notably, entanglement persists even at temperatures up to $T=1\,\mathrm{K}$. The amount of bipartite entanglement is enhanced when the Barnett shift changes from $\hat{\Delta}_{B}=0$ (non-spinning YIG sphere) to $\hat{\Delta}_{B}=0.3\,\omega_{\rm{b_1}}$.  
 	\begin{figure}[H]
 	\centering
 	{\label{figure44c}\includegraphics[scale=0.61]{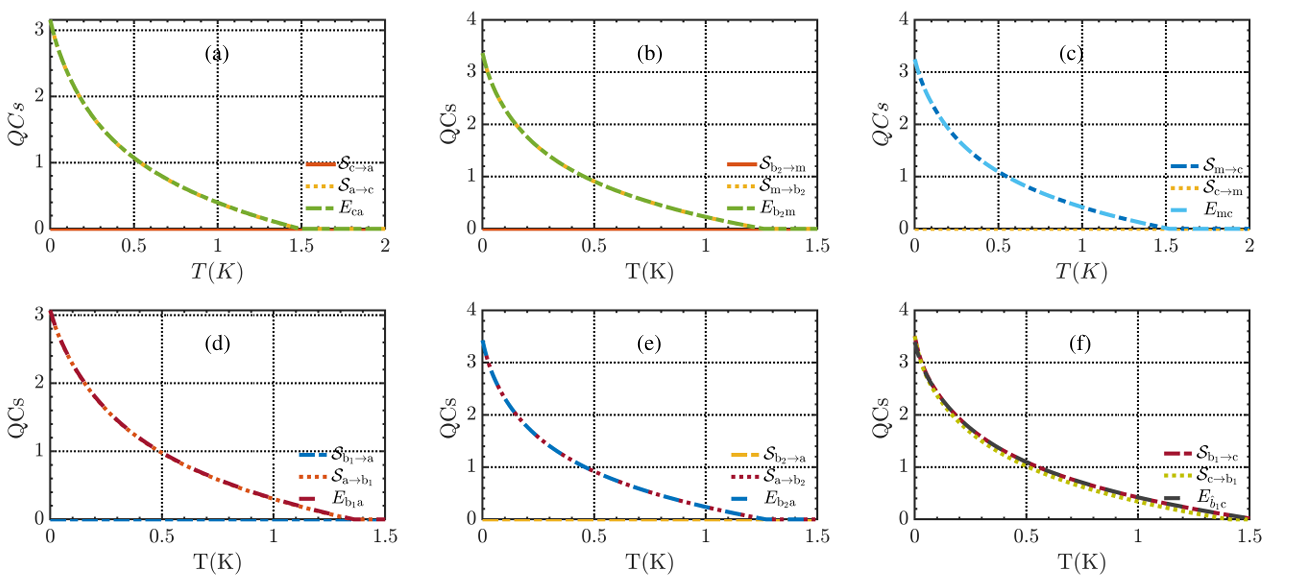}}
 	
 	\caption{ The Qcs ( steerings $\mathcal{S}_{  {s } \rightarrow s'}$,  $\mathcal{S}_{ s' \rightarrow s}$ and entanglement $E_{ss' }$) between the six nondirectly  coupled modes, vary in respect to the temperature $T$ variation. For $ {\mathcal{L}}=0.9$ and $\hat{\Delta}_{B}=0.2 \omega_{\rm{b}_{1}}$.}  
 	\label{P}
 \end{figure}

 \begin{figure}[h]
 	\centering
 	{\label{figure7a}\includegraphics[scale=0.69]{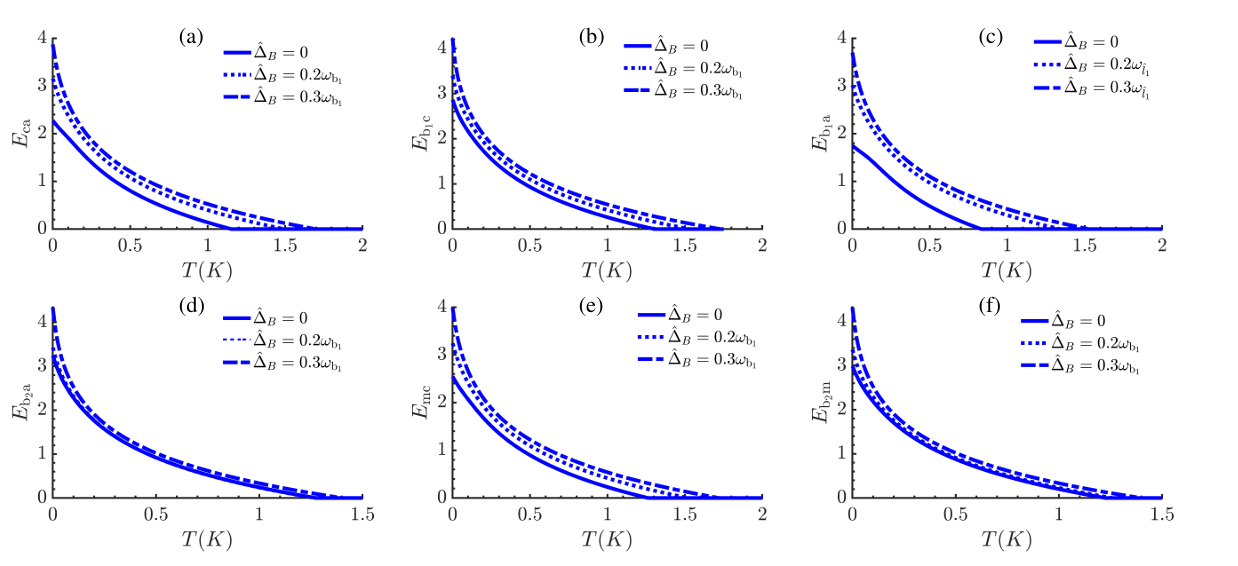}}
 	
 	\caption{Entanglement versus temperature for different Barnett shift ($\hat{\Delta}_{B}$) values. The subplots show entanglement ($E_{ss'}$) between: (a) microwave and optical modes ($E_{\rm{ca}}$), (b) mechanical mode $\rm{b_1}$ and optical mode ($E_{\rm{b_1c}}$), (c) mechanical mode $\rm{b_2}$ and microwave mode ($E_{\rm{b_2a}}$), (d) magnon and optical modes ($E_{\rm{mc}}$), (e) mechanical mode $\rm{b_1}$ and optical mode ($E_{\rm{b_1c}}$), and (f) mechanical mode $\rm{b_2}$ and magnon mode ($E_{\rm{b_2m}}$).}
 	\label{AF1}
 \end{figure} 

\begin{figure}[H]
	\centering
	{\label{figure7a}\includegraphics[scale=0.65]{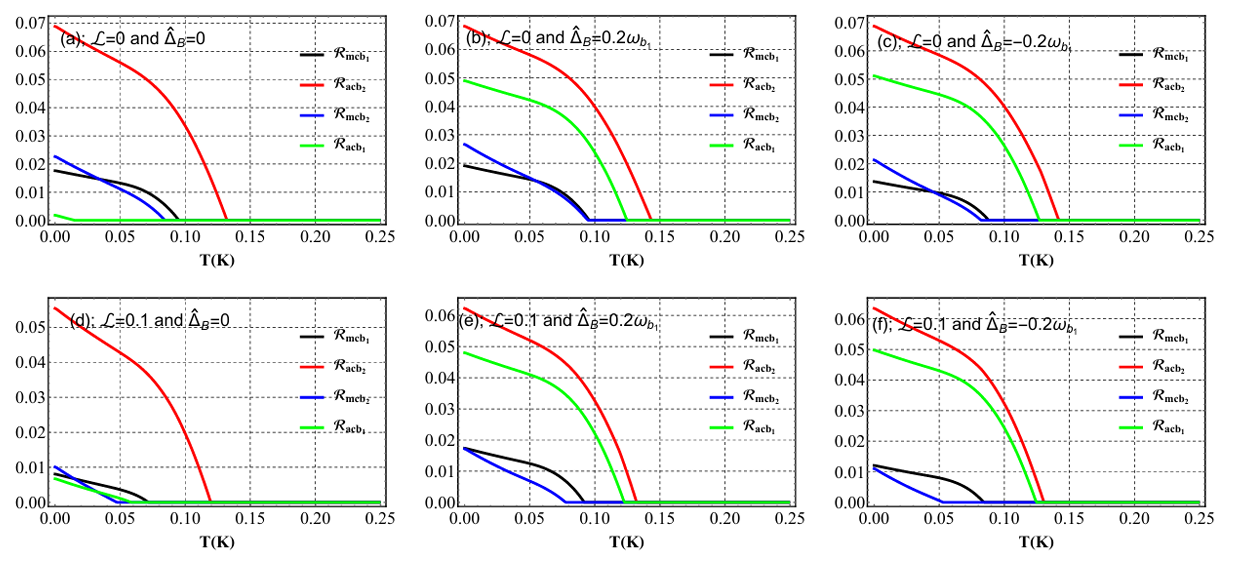}}
	
\caption{Tripartite entanglement measures $\mathcal{R}_{\rm{mcb_1}}$, $\mathcal{R}_{\rm{acb_2}}$, $\mathcal{R}_{\rm{mcb_2}}$, and $\mathcal{R}_{\rm{acb_1}}$ are all plotted as a function of temperature. These plots show results for various values of the Barnett shift ($\hat{\Delta}_{B}$) and the reflectivity parameter ($\mathcal{L}$). All other parameters are the same as those used in Fig.~\ref{r2}.}

	\label{AF11}
\end{figure}

In Fig. \ref{AF11}, we represent   the tripartite entanglement as a function of temperature for different reflectivity parameter ($\mathcal{L}$) and Barnett shifts ($\hat{\Delta}_{B}$). We observe that when only the BE tool is used, the level of tripartite entanglement increases for all modes and becomes more robust against thermal effects, confirming our previous remarks in Fig. \ref{P}. Tripartite entanglement for $\mathcal{R}_{\rm{mcb_1}}$, $\mathcal{R}_{\rm{acb_1}}$, and $\mathcal{R}_{\rm{mcb_2}}$ exhibits asymmetric behavior when the YIG sphere's rotation direction changes, as seen in Figs. \ref{AF11}(b) and (c). When both BE and FB tools are combined (Figs. \ref{AF11}(e) and (f)), the level of tripartite entanglement for $\mathcal{R}_{\rm{acb_1}}$ can be enhanced compared to when both tools are absent (Fig. \ref{AF11}(a)) and when only FB is incorporated (Fig. \ref{AF11}(d)).
  
\begin{figure}[H]
	\centering		
	{\label{figure9b}\includegraphics[scale=0.65]{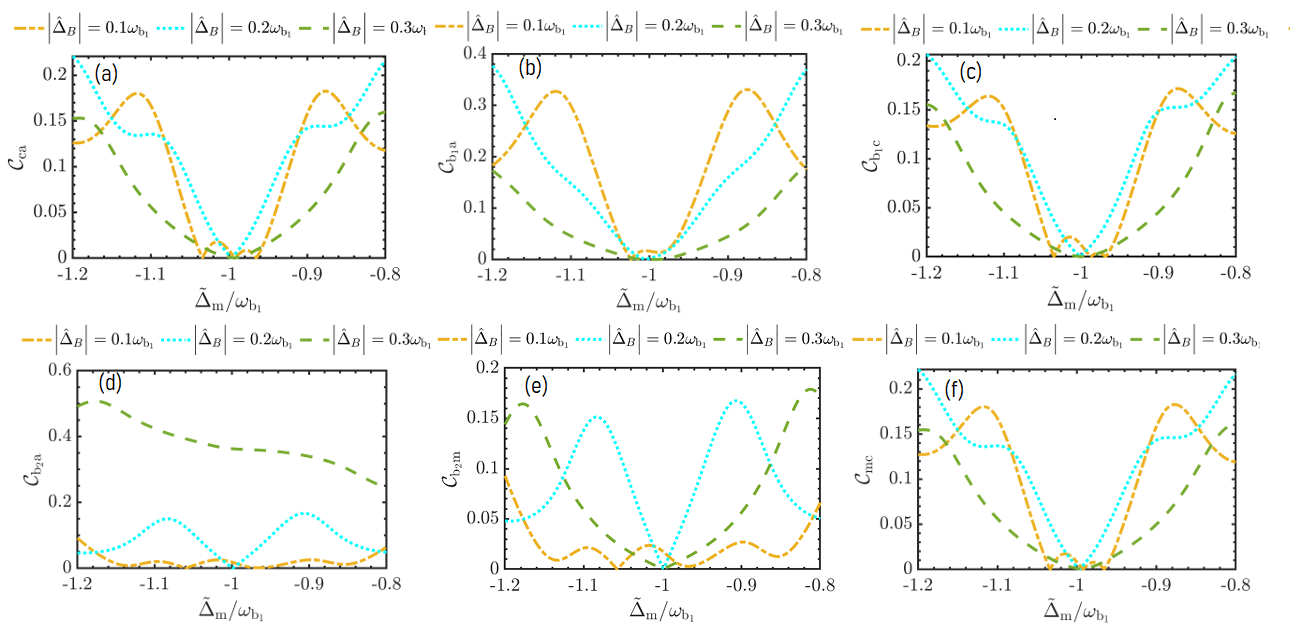}}

	\caption{The nonreciprocity $\mathcal{C}_{ss'}$ measures as  a function of the detuning of the magnon mode $\tilde{\Delta}_{\rm{m}}$, using ${\mathcal{L}}=0.6$. The others parameters are the same used on the previous figures.}
	\label{rec1}
\end{figure}

\begin{figure}[H]
\centering
{\label{figure9b}\includegraphics[scale=0.65]{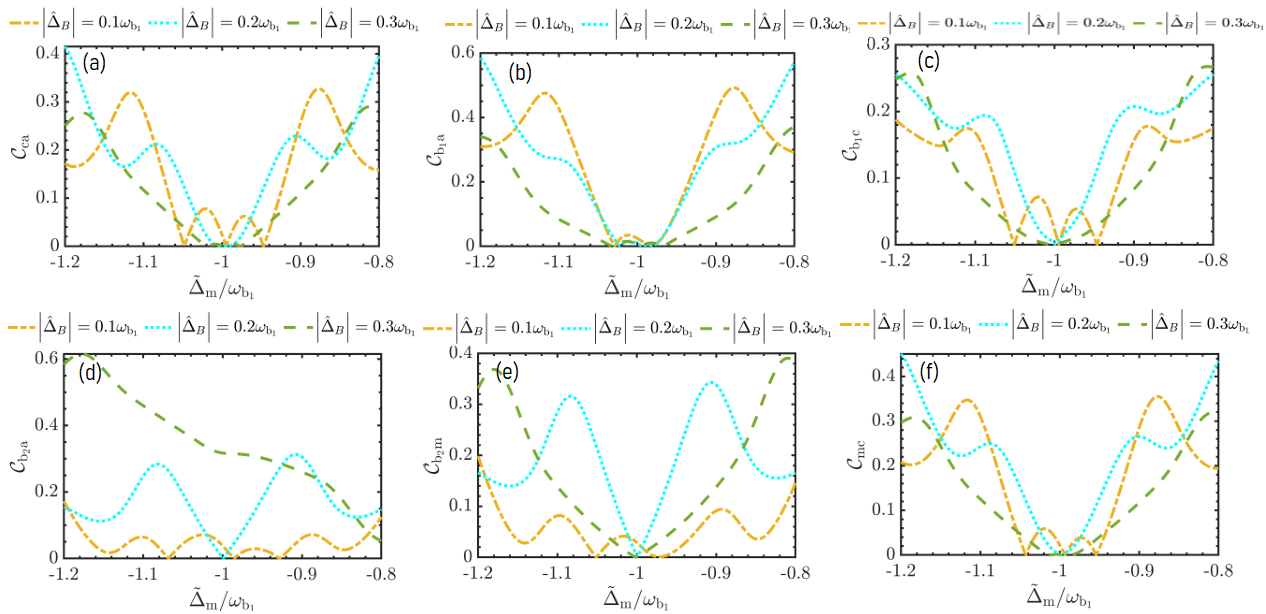}}
			
			\caption{The  nonreciprocity $\mathcal{C}_{ss'}$ measures as a function of the detuning of the magnon mode $\tilde{\Delta}_{\rm{m}}$, with  ${\mathcal{L}}=0.9$. The others parameters are the same used on the previous curves.}
			\label{NR}
		\end{figure}

To quantitatively characterize nonreciprocal entanglement, we introduce the bidirectional contrast ratio ( $\mathcal{C}$ ) (ranging from 0 to 1) as a metric for evaluating both bipartite and tripartite entanglement in nonreciprocal regimes.
\begin{equation}
	\begin{aligned}
		C_{s s^{\prime}} &= \frac{\left| E_{s s^{\prime}}(\hat{\Delta}_B > 0) - E_{s s^{\prime}}(\hat{\Delta}_B < 0) \right|}{E_{s s^{\prime}}(\hat{\Delta}_B > 0) + E_{s s^{\prime}}(\hat{\Delta}_B < 0)}, \\
		C_{\mathcal{R}} &= \frac{\left| \mathcal{R}_\tau^{\min}(\hat{\Delta}_B > 0) - \mathcal{R}_\tau^{\min}(\hat{\Delta}_B < 0) \right|}{\mathcal{R}_\tau^{\min}(\hat{\Delta}_B > 0) + \mathcal{R}_\tau^{\min}(\hat{\Delta}_B < 0)}.
	\end{aligned}
\end{equation}

when \( C_{s s^{\prime}}(C_{\mathcal{R}}) = 1 \), the system displays perfect nonreciprocity, whereas \( \mathcal{C}_{s s^{\prime}} (C_{\mathcal{R}})  = 0 \) signifies the absence of nonreciprocal  in bipartite (tripartite) entanglements.
To illustrate this behavior, Figs. \ref{rec1} and \ref{NR} present the variation of the bidirectional contrast ratio $C$ as a function of the normalized magnon detuning $\tilde{\Delta}_{\rm{m}}/ \omega_{\rm{b}_1} $ for different values of $\hat{\Delta}_B$. This is shown for $\mathcal{L}=0.6$ (Fig. \ref{rec1}) and for $\mathcal{L}=0.9$ (Fig. \ref{figure9}). The figures clearly demonstrate that nonreciprocity in bipartite entanglements can be effectively controlled, turned on, or turned off by tuning the magnon mode detuning and adjusting the angular frequencies. A maximum value of nonreciprocal entanglement can be achieved by carefully setting the angular frequencies. Generally, across multiple bipartitions, a higher angular frequency proves advantageous for achieving stronger entanglement nonreciprocity. As previously discussed, keeping the YIG sphere rotating in a fixed direction while applying a magnetic field along the $+z$ (or $-z$) axis induces a positive (or negative) frequency shift $\hat{\Delta}_B$ due to the BE \cite{97be, 99be}, which is analogous to the Sagnac effect \cite{71be}. As a result, the BE facilitates the emergence of nonreciprocal bipartite entanglement. The nonreciprocity degree of entanglement can also be improved by an appropriate choice of the reflectivity parameter $\mathcal{L}$. We observe that increasing the reflectivity parameter from 0.6 to 0.9 enhances the degree of nonreciprocity, thereby demonstrating the relationship between this parameter and the control of entanglement nonreciprocity in the proposed system.

\begin{figure}[H]
\centering
{\label{figure9b}\includegraphics[scale=0.60]{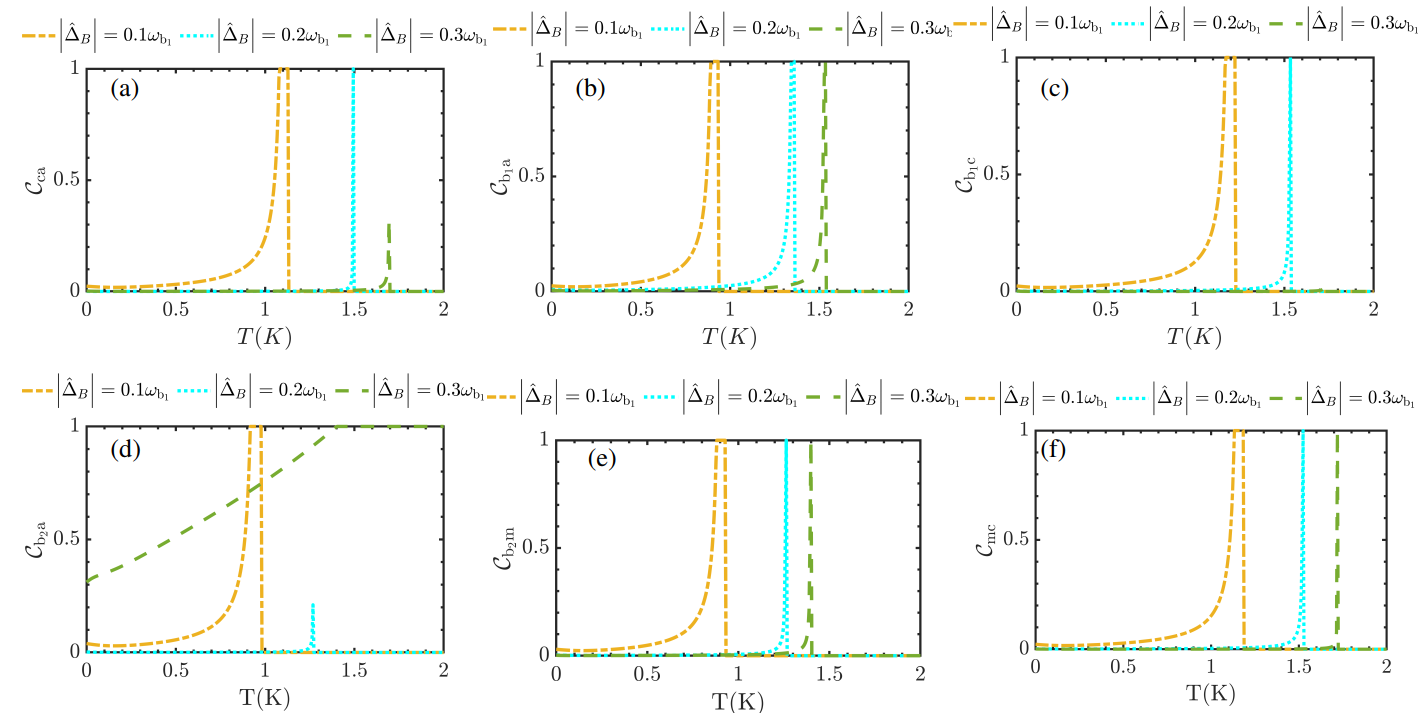}}
\caption{The nonreciprocity $\mathcal{C}_{ss'}$ measures as function of the temperature $T$ with ${\mathcal{L}}=0.9$. The others parameters are the same used on the text.}
\label{rec11}
\end{figure}

\begin{figure}[H]
			\centering
			{\label{figure9a}\includegraphics[scale=0.65]{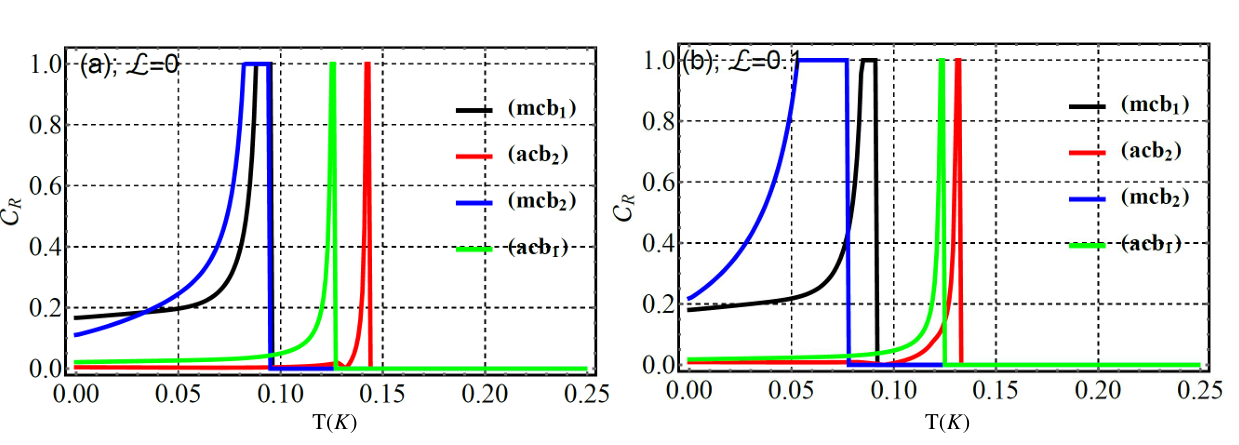}}
			\caption{The tripartite nonreciprocity $\mathcal{C}_{R}$ measures as function of the temperature $T$, for ${\mathcal{L}}=0$ (a) and for ${\mathcal{L}}=0.1$ (b). The others parameters are the same used in the previous curves.}
			\label{figure9}
		\end{figure}

Then, the Fig. \ref{rec11} illustrates the variations of the bipartite nonreciprocity of entanglement as a function of the temperature. The figure shows clearly how the temperature value affects the nonreciprocal entanglement, indicating that the entanglement of all the pairs can be nonreciprocally enhanced by the temperature effect \cite{105}. The nonreciprocity of entanglement initially increases gradually with temperature $T$. As $T$ continues to rise, a sharp enhancement emerges, leading to ideal nonreciprocal entanglement across many bipartitions. A bidirectional contrast ratio of $C_{ss'} = 1$ signifies perfect nonreciprocity, meaning that entanglement exists ($E_{ss'} > 0$) in one direction of rotation but is entirely suppressed  $E_{ss'} = 0$ in the opposite direction. 
Finally, Fig. \ref{figure9} explores the tripartite nonreciprocity with respect to temperature variations when $\mathcal{L}=0$ (Fig. \ref{figure9} (a)) and when $\mathcal{L}=0.1$ (Fig. \ref{figure9} (b)). We see that by carefully adjusting the reflectivity parameter, we can increase the range in which the nonreciprocity of the tripartite entanglement gets maximum. The contrast ratio can be set between $0$ and $1$ (ideal), showing how the reflectivity parameter $\mathcal{L}$ relates to the tripartite nonreciprocity. The nonreciprocity of the entanglements increases gradually, first with $T$. By further increasing $T$, a sharp rise occurs, and ideal nonreciprocal entanglements can be achieved across all four combinations. The case when $C_\mathcal{R}=1$ corresponds to the perfect nonreciprocity of entanglement, i.e., the entanglement is present ($\mathcal{R} > 0$ ) in one direction of rotation  but completely absent $\mathcal{R}=0$ in the reverse direction. 		

\subsection{EXPERIMENTAL FEASIBILITY}

In this subsection, we briefly discuss the experimental feasibility of the proposed scheme. Based on recent advances, we consider the following experimentally achievable parameters: $\omega_{\rm{a}, \rm{m}} / 2\pi = 10\ \mathrm{GHz}$, $\omega_{\rm{b}_1} / 2\pi = 20.15\ \mathrm{MHz}$, $\omega_{\rm{b}_2} / 2\pi = 20.11\ \mathrm{MHz}$, optical cavity resonance wavelength $\lambda_c = 1550\ \mathrm{nm}$, damping rates $\gamma_{\rm{a}, \rm{m}, \rm{c}} / 2\pi = 1\ \mathrm{MHz}$ and $\gamma_{\rm{b}_1, \rm{b}_2} / 2\pi = 100\ \mathrm{Hz}$, effective coupling rates ${\hat{\mathcal{D}}}_m / 2\pi = 0.7\ \mathrm{MHz}$ and $\hat{\mathcal{D}}_{\rm{c}} / 2\pi = 2.7\ \mathrm{MHz}$, and a temperature $T = 10\ \mathrm{mK}$. In experimental setups, a YIG sphere can be rotated by attaching it to a thin rod and mounting the assembly onto a rotor or turbine, similar to techniques used with rotating optical spherical resonators \cite{exp1}. Remarkably, rotational speeds in the gigahertz range have been demonstrated using levitated nanoparticles \cite{62exp}. However, electrically driven systems may generate considerable electromagnetic background noise. As an alternative, the YIG sphere can be mounted on an air turbine and driven by airflow to mitigate this issue \cite{exp3}.
Additionally, the Barnett shift is directly influenced by the rotational angular velocity of the YIG sphere. Recent experimental results demonstrate that this velocity can reach the GHz range using a levitated crystal \cite{62exp,63exp}. While such high rotational speeds can lead to temperature increases, this issue can be mitigated by operating the system in a low-temperature vacuum environment. The  feedback control systems  shown strong tunability and proven feasible in experimental settings \cite{exp8,exp10}.

\section{Conclusions}  \label{conc}

In this work, we proposed a scheme based on a practical experimental system to generate and control asymmetric steering and stationary bipartite and tripartite entanglement by exploring the FB tool and the BE. Increasing the reflectivity parameter $\mathcal{L}$ improves the amount of QCs. By combining both FB and BE, we can significantly enhance the manipulation and control of entanglement and asymmetric EPR steering. The BE tool contributes to generating both nonreciprocal bipartite and tripartite entanglement. We also observed the emergence of entanglement between the phonon $\rm{b}_1$ and the optical mode when using FB alone, which is further enhanced by the combined use of FB and BE. Overall, BE plays a crucial role in making the QCS more robust against thermal effects. Nonreciprocity in both bipartite and tripartite entanglement can be achieved by appropriately tuning the system parameters.

\end{document}